\documentclass[%
 reprint,
 amsmath,amssymb,
 aps,
]{revtex4-1}

\usepackage{graphicx}% Include figure files
\usepackage{dcolumn}% Align table columns on decimal point
\usepackage{bm}% bold math
\begin{document}

\preprint{APS/123-QED}

\title{Presupernova neutrino events relating to the final evolution of massive stars}% Force line breaks with \\

\author{Takashi Yoshida$^{1}$}
\email{tyoshida@astron.s.u-tokyo.ac.jp}

\author{Koh Takahashi$^{1}$}%

\author{Hideyuki Umeda$^{1}$}

\author{Koji Ishidoshiro$^{2}$}
\affiliation{%
$^{1}$Department of Astronomy, Graduate School of Science, University of Tokyo, Tokyo 113-0033, Japan\\
$^{2}$Research Center for Neutrino Science, Tohoku University, Miyagi 980-8578, Japan}%
%\affiliation{
%$^{2}$Research Center for Neutrino Science, Tohoku University, Miyagi 980-8578, Japan}%


%\affiliation{
%\affiliation{
%\affiliation{%


\date{\today}% It is always \today, today,

\begin{abstract}
When a supernova explosion occurs in neighbors around hundreds pc, 
current and future neutrino detectors are expected to observe neutrinos
from the presupernova star before the explosion.
We show a possibility for obtaining the evidence for burning processes
 in the central region of presupernova stars though the observations of neutrino signals 
by current and future neutrino detectors such as KamLAND, JUNO, and Hyper-Kamiokande.
We also investigate supernova alarms using neutrinos from presupernova stars in neighbors.
If a supernova explodes at $\sim 200$ pc, future 20 kton size liquid scintillation detectors are 
expected to observe hundreds neutrino events.
We also propose a possibility of the detection of neutrino events 
by Gd-loaded Hyper-Kamiokande using delayed $\gamma$-ray signals.
These detectors could observe detailed time variation of neutrino events.
The neutrino emission rate increases by the core contraction in the final evolution stage.
However, the O and Si shell burnings suppress the neutrino emission for a moment.
The observed decrease in the neutrino event rate before hours to the explosion is 
possibly evidence for the shell burnings.
The observations of detailed time evolution of presupernova neutrino events  
could reveal properties of burning processes in the central region of presupernova stars.
\end{abstract}

%\keywords{neutrinos --- stars: evolution --- stars: massive --- supernovae: general ---}%Use showkeys class option if keyword
\maketitle

%\tableofcontents

\section{Introduction}

Neutrinos emitted from core-collapse (CC) supernovae (SNe) give the information of the central interior 
of the collapsing core in an evolved massive star.
Kamiokande and Irvine-Michigan-Brookhaven (IMB) experiment observed twelve and eight neutrino
events in the explosion SN 1987A \cite{Hirata87,Bionta87}.
These observations confirmed basic characteristics of current SN models and neutron
star formation.
If a SN explodes at the Galactic center, thousands of neutrino events will be detected by
Super-Kamiokande (e.g., \cite{AbeK16}) and hundreds events will be by KamLAND and 
other neutrino detectors.

There are also evolved massive stars such as red supergiants and Wolf--Rayet (WR) stars in the
distance of hundreds pc.
One famous example is Betelgeuse, a red supergiant at $197 \pm 45$ pc and its initial mass
is considered to be $\sim$17 $M_\odot$ \cite{Harper08}. 
If these massive stars become SNe, millions neutrinos will be detected by Super-Kamiokande.
On the other hand, neutrinos are also emitted from SN progenitors before the SN explosion.
Indeed, neutrino emission is the most efficient cooling process after the He burning.
The neutrino luminosity during the Si burning becomes the order of $10^{47}$ erg s$^{-1}$.

Recently, properties of neutrino spectra emitted by various neutrino emission processes in evolved
massive stars have been investigated: pair neutrino process \cite{Odrzywolek04, Misiaszek06}, 
plasma neutrinos \cite{Odrzywolek07},  and weak interactions of nuclei such as electron captures 
in nuclear statistical equilibrium \cite{Odrzywolek09, Odrzywolek10, Patton15}.
The time evolution of the neutrino emission rate has been investigated \cite{Odrzywolek10}.
Neutrinos from presupernova (preSN) stars are less energetic and less luminous than
SN neutrinos.
Nevertheless, since liquid scintillation neutrino detectors such as KamLAND have low threshold energy, 
they are expected to detect preSN neutrinos if a SN explodes in neighbors.
Depending on the stellar models and the distance, tens $\sim 100$ neutrino events from preSN stars 
of nearby SNe have been expected to be detected by KamLAND \cite{Odrzywolek07, Kato15, Asakura16}. 
The detectability of the neutrinos from the progenitor of an electron capture SN is also investigated 
\cite{Kato15}.
A SN alarm using the detection of neutrinos from a preSN star could be possible by KamLAND 
within a few to tens of hours before a SN explosion \cite{Asakura16}.

\begin{table*}
\caption{Properties of the 12, 15, and 20 $M_\odot$ models.
$L_{\rm f}$ is the luminosity at the last step.
}
\begin{center}
\begin{tabular}{ccccccccc}
\tableline\tableline
$M_{\rm init}$ ($M_\odot$) & $M_{\rm fin}$ ($M_\odot$) & $M_{\rm He}$ ($M_\odot$) & 
$M_{\rm CO}$ ($M_\odot$) & $M_{\rm Si}$ ($M_\odot$) & $M_{\rm Fe}$ ($M_\odot$) & 
$t_{\rm Si-b}$ (days) & $t_{\rm col}$ (hours) & $\log (L_{\rm f}/L_\odot)$ \\
\tableline
12 & 10.6  & 3.52 & 1.82  & 1.53 & 1.38 & 8.6 & 30.0 & 4.75 \\
15 & 12.3  & 4.66 & 2.74  & 1.65 & 1.50 & 4.4 & 21.4 & 4.98 \\
20 & 14.3  & 6.86 & 4.64  & 2.11 & 1.44 & 1.1 & 10.4 & 5.28 \\
\tableline\tableline
\end{tabular}
\end{center}
\label{tab:stars}
\end{table*}

We still have no ways to observe directly deep interior of evolved stars.
The advanced stellar evolution of massive stars has been understood through findings of 
theoretical studies on stellar models 
(e.g., \cite{Weaver78, Nomoto88, Limongi00, Woosley02, Hirschi04, Umeda12}).
At the final stage, the Si core burning, the Si shell burning as well as the
O shell burning proceeds complicatedly in the central region.
The Si core burning proceeds for a few days to one week to form an iron core.
The iron core grows up with the Si shell burning and collapses for hours.
The structure change by burning processes and the collapse affects properties
of neutrinos emitted from a preSN star.
If the evolution of detailed neutrino emissivity or neutrino spectra is observed by current
or future neutrino detectors, these observational data will give constraints to the
core structure and burning processes of preSN stars.
KamLAND is a one kton size liquid scintillation neutrino detector \cite{Asakura16}.
Twenty kton size liquid scintillation detectors such as JUNO \cite{An15} 
and RENO-50 \cite{Seo15} are planned.
A hundreds kton size water Cherenkov detector Hyper-Kamiokande is also planned \cite{AbeK11a}.
Thus, it is important to investigate the relations between detailed neutrino properties from preSN stars 
and the final evolution of the stars.

The purposes of this study are to estimate preSN neutrino events observed by current and future
neutrino detectors such as KamLAND, JUNO, and Hyper-Kamiokande and to find the relations
between detailed time variation of the neutrino events observed by these detectors and 
burning processes occurred in the central region of preSN stars.
We also estimate the SN alarms using the observations of preSN neutrinos.
We investigate the time evolution of the spectra of neutrinos emitted through
pair-neutrino process from the Si burning of 12, 15, and 20 $M_\odot$ stars
taking into account the stellar structure.
Then, we estimate preSN neutrino events observed by current and future neutrino detectors.
We discuss the relation between the evolution of preSN neutrinos and burning processes 
after the Si core burning.

We will organize this article as follows.
We present the stellar evolution model and properties of neutrinos produced by pair neutrino process
in Sec. II.
In Sec. III, we show properties of neutrino spectra and the corresponding internal structure 
during the evolution from the Si burning in the 15 $M_\odot$ model.
We also show the stellar mass dependence of the neutrino properties.
In Sec. IV, we estimate the neutrino events by current and future neutrino detectors assuming
the distance to a preSN star of 200 pc.
We discuss burning processes suggested by the evolution of preSN neutrino events.
The SN alarm using preSN neutrinos by KamLAND and JUNO is also discussed.
In Sec. V, we discuss preSN neutrino events from Betelgeuse considering the uncertainty of the distance.
We also discuss the relation between the preSN neutrino events and parameters 
characterizing SN explosion.
We give conclusions in Sec. VI.

\section{Calculation method}

\subsection{Stellar evolution model}

We calculate the evolution of massive stars with the initial mass of 12, 15, and 20 $M_\odot$ and the
solar metallicity from the zero-age main sequence to the onset of the core-collapse when the central
temperature reaches $\sim 10^{9.8}$ K.
The main input physics of the stellar evolution is written in \cite{Takahashi16}.
We use the nuclear reaction network of 300 species of nuclei \cite{Yoshida14}.
The stellar mass reduces by the mass loss process during the evolution up to the C burning.
The mass loss rate for the main-sequence stage and red supergiant stage is adopted from 
\cite{Vink01, Nieuwenhuijzen90}.
In the stellar evolution models, the neutrino energy loss rate by pair, photo, plasma, 
Bremsstrahlung, and recombination neutrino processes are included.
These rates are calculated using the approximated formula in \cite{Itoh96}.
We also calculate the neutrino energy loss by weak interactions of nuclei in calculating the abundance 
evolution using tables in \cite{Fuller82, Oda94, Langanke01}.

\begin{figure}[b]
\includegraphics[width=6cm,angle=270]{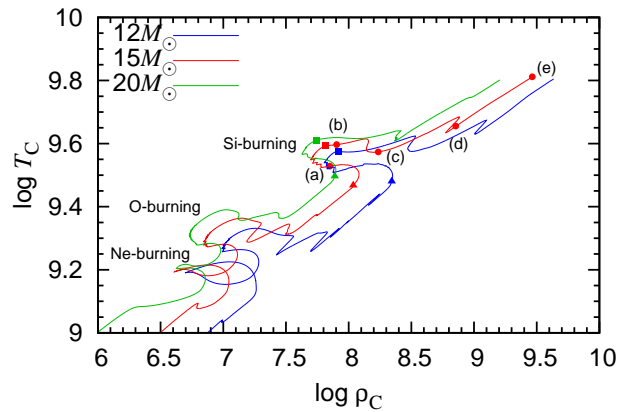}
\caption{
The evolution of the central density and temperature in the 12 (blue line), 15 (red line), and 
20 (green line) $M_\odot$ models after the C burning.
Triangles and rectangles indicate the ignition and termination of the Si core burning, respectively.
Mass fraction distribution and neutrino properties at the stages assigned by circles 
with symbols (a)--(e) in the 15 $M_\odot$ model are explained in Sec. III A 2.
\label{fig:rhoctc}
}
\end{figure}

\begin{table*}
\caption{Stellar properties and average neutrino energy at five different stages of the 15 $M_\odot$ model.
}
\begin{center}
\begin{tabular}{ccccccccc}
\tableline\tableline
Stage & Time to collapse  & $\log T_{\rm C}$ & $\log \rho_{\rm C}$ 
& $\langle \varepsilon_{\nu_{e}} \rangle$ (MeV) & $\langle \varepsilon_{\bar{\nu}_{e}} \rangle$ (MeV) & 
$\langle \varepsilon_{\nu_{\mu,\tau}} \rangle$ (MeV) & 
$\langle \varepsilon_{\bar{\nu}_{\mu,\tau}} \rangle$ (MeV) & Main burning \\
\tableline
(a) & 3.61 days  & 9.53 & 7.85  & 1.19 & 1.09 & 1.28 & 1.24 & Si core burning \\
(b) & 20.1 hours  & 9.60 & 7.90  & 1.45 & 1.33 & 1.50 & 1.47 & After Si core burning \\
(c) & 9.78 hours  & 9.57 & 8.24  & 1.27 & 1.18 & 1.36 & 1.33 & O shell burning \\
(d) & 23.8 minutes & 9.66 & 8.85  & 1.51 & 1.42 & 1.58 & 1.56 & Si shell burning \\
(e) & 0.00 second  & 9.81 & 9.46  & 2.10 & 1.94 & 2.14 & 2.11 & Si shell burning \\
\tableline\tableline
\end{tabular}
\end{center}
\label{tab:5pt}
\end{table*}

The final mass, the masses of He, CO, Si, and Fe cores are listed in Table \ref{tab:stars}.
We set the outer boundaries of the He, CO, and Si cores as the outermost mass coordinates where
the mass fractions of H, He, and O are smaller than 0.1, respectively.
For Fe-core mass, we set the boundary as the outermost mass coordinate 
where the mass fraction of Fe-peak elements, denoted as ^^ ^^ Fe," the elements with 
atomic number $Z \ge 22$, is larger than 0.5.
In these models, the masses of the He, CO, and Si cores monotonically increase with the initial mass.
We also list the periods between the ignition and termination of the Si core burning $t_{\rm Si-b}$
and from the termination of the Si core burning until the last step of the calculations $t_{\rm col}$, 
and the luminosity at the last step.
The period of the Si-burning is one day to one week.
It decreases with increasing the initial stellar mass.

We show the evolution of the central temperature and density after the C burning to the core-collapse
in Fig. \ref{fig:rhoctc}.
We also list some stellar properties at five different stages (a)--(e) of the 15 $M_\odot$ model 
in Table \ref{tab:5pt}.
During most of the period of the Si core burning, assigned by triangle and rectangle 
in Fig. \ref{fig:rhoctc}, the central temperature slightly raises and the 
central density decreases.
Stage (a) is located during the expansion by the Si core burning.
When the Si core burning weakens, the density turns to raise again.
Stage (b) is located just after the termination of the Si core burning.
Then, the central temperature decreases once at around $\log T_{\rm C} \sim 9.6$ 
close to stage (c) in Fig. \ref{fig:rhoctc}.
At this time, the O shell burning starts at $M_{\rm r} \sim 1.4 M_\odot$.
There are also dips around $\log T_{\rm C} \sim 9.65$ in the 12 and 15 $M_\odot$ models
[see stage (d) for the 15 $M_\odot$ model].
The main burning process changes to the Si shell burning in $M_{r} \gtrsim 1 M_\odot$.
Stage (e) is located at the last step of the calculation.

\subsection{Neutrino emission by pair neutrino process}

Pair neutrino process is the neutrino emission process through the pair annihilation of electrons
and positrons.
This process is a dominant neutrino emission process during most advanced stages of massive stars
(e.g., \cite{Itoh96}).
Here, we evaluate the spectra of neutrinos and antineutrinos emitted through pair neutrino process.
The neutrino emission rate has been evaluated in \cite{Dicus72} 
(see also \cite{Yakovlev01}).
We use the description of the emission rate of a given flavor of neutrinos
per unit volume as
\begin{eqnarray}
r(\varepsilon_{\nu}, \varepsilon_{\bar{\nu}}) && = \\
&& \frac{c}{16 (2\pi \hbar)^{12}} \int f_{\rm e^{-}} f_{\rm e^{+}}
(2\pi)^4 \delta^4(p_{\rm e^{-}}+p_{\rm e^{+}}-p_{\nu}-p_{\bar{\nu}}) \nonumber \\
&& \times \frac{|M|^2}{\varepsilon_{\rm e^{-}} \varepsilon_{\rm e^{+}} \varepsilon_{\nu} \varepsilon_{\bar{\nu}} }
d^3 p_{\rm e^{-}} d^3 p_{\rm e^{+}} d\Omega_{\nu} d\Omega_{\bar{\nu}} , \nonumber
\end{eqnarray}
where $\varepsilon_\nu$ and $\varepsilon_{\bar{\nu}}$ are the energies of neutrinos
and antineutrinos of a given flavor, $\hbar$ is the reduced Planck constant, $c$ is the light speed, 
$f_i$, $p_i$, $\varepsilon_i$, and $\Omega_i$ are the Fermi-Dirac distribution, 
four-dimensional momentum, energy, and solid angle of particle $i$.
The matrix element $|M|^2$ is written as
\begin{eqnarray}
|M|^2 &=& 16 G_{\rm F}^2 (\hbar c)^2 \{
 (C_{\rm A} - C_{\rm V})^2 (p_{\rm e^{-}} \cdot p_{\nu}) (p_{\rm e^{+}} \cdot p_{\bar{\nu}}) \\
&& + (C_{\rm A} + C_{\rm V})^2 (p_{\rm e^{+}} \cdot p_{\nu}) (p_{\rm e^{-}} \cdot p_{\bar{\nu}})  \nonumber \\
&& + m_e^2 c^4 (C_{\rm A}^2 + C_{\rm V}^2) (p_{\nu} \cdot p_{\bar{\nu}}) \},  \nonumber
\end{eqnarray}
where $G_{\rm F}$ is the Fermi-coupling constant, $C_{\rm V}$ and $C_{\rm A}$ are 
the vector and axial-vector coupling constants, respectively, $m_e$ is the electron mass.
The Fermi-Dirac distribution function depends on the temperature $T$ and the multiple
of  the density $\rho$ and the electron mole fraction $Y_e$.
The value of $C_{\rm V}$ is set to be $1/2+2\sin ^2\theta_{\rm W}$ and $1/2-2\sin ^2\theta_{\rm W}$
for $\nu_e\bar{\nu}_e$ pair and $\nu_{\mu}\bar{\nu}_{\mu}$ and $\nu_{\tau}\bar{\nu}_{\tau}$ pairs,
respectively, where $\theta_{\rm W}$ is the Weinberg angle and we set 
$\sin ^2\theta_{\rm W} = 0.23126$ \cite{Olive14}.
The value of $C_{\rm A}$ is set to be 1/2.
The $\nu_e\bar{\nu}_e$ pair is produced through neutral-current and charged-current
processes.
The pairs of $\nu_{\mu}\bar{\nu}_{\mu}$ and $\nu_{\tau}\bar{\nu}_{\tau}$ are produced through
neutral-current process.

\begin{figure}[b]
\includegraphics[width=6cm,angle=270]{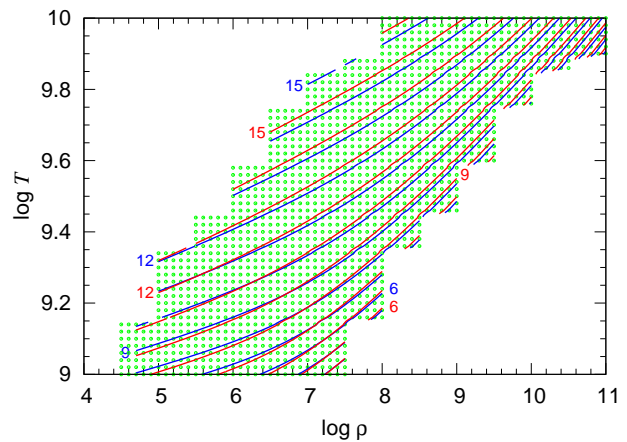}
\caption{
Contours of the energy loss rate in the unit mass $d\epsilon_{\nu}/dt$
by $\nu_{e}\bar{\nu}_{e}$ pairs (red lines) and 
$\nu_{x}\bar{\nu}_{x}$ pairs where $x$=$\mu$ or $\tau$ (blue lines)
on the $\log \rho$ -- $\log T$ plane.
The $Y_e$ value is assumed to be 0.5.
The numbers attached to lines indicate $\log (d\epsilon_{\nu}/dt)$.
The neutrino spectra at the points assigned by small green circles on this plane are calculated
in this study.
\label{fig:qnucnt}
}
\end{figure}

\begin{figure}
\includegraphics[width=6cm,angle=270]{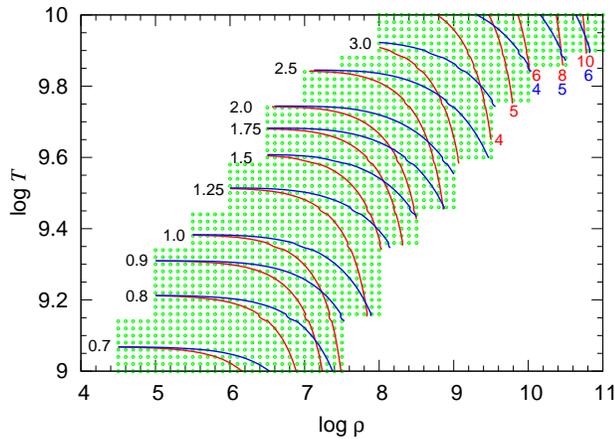}
\caption{
Contours of the average energy of $\nu_{e}$ (red lines) and $\bar{\nu}_{e}$ (blue lines) 
emitted by pair neutrino process in units of MeV on the $\log \rho$ -- $\log T$ plane.
The $Y_e$ value is assumed to be 0.5.
Black numbers indicate the energy value of both of $\nu_e$ and $\bar{\nu}_e$.
Red and blue numbers indicate the values of the corresponding contours of $\nu_e$ and $\bar{\nu}_e$,
respectively.
\label{fig:eavcnt}
}
\end{figure}

In this study, we calculate the neutrino spectra at 1581 points on the plain of $\log \rho$ and 
$\log T$ assigned by small green circles in Figs. \ref{fig:qnucnt} and \ref{fig:eavcnt} and assuming
$Y_e = 0.5$.
Integration of the phase space has been performed using Monte-Carlo method.
This region covers the ranges of the density and temperature in the stellar evolution models
where the temperature is larger than $1 \times 10^9$ K.
We evaluate the neutrino spectra at each time and each mass coordinate by interpolating 
the spectra of four neighboring points on the $\log \rho Y_{e}$ and $\log T$ plain.

Figure \ref{fig:qnucnt} shows contours of the energy loss rate $d\epsilon_{\nu}/dt$ 
(erg s$^{-1}$ g$^{-1}$) by pair neutrino process.
The energy loss rate increases with temperature and decreases with increase in density.
When the density is high, the electron degeneracy becomes large and positron number 
decreases.
Thus, the rate of pair neutrino process and the energy loss rate decreases. 
Figure \ref{fig:eavcnt} shows contours of the average energies of $\nu_e$ and $\bar{\nu}_e$.
At a given density, the average energies of $\nu_e$ and $\bar{\nu}_e$ increase with temperature.
The average energy of $\nu_e$ is larger than that of $\bar{\nu}_e$.
When the electron degeneracy is small, the difference of the average $\nu_e$ energy and
$\bar{\nu}_e$ energy is small.
The difference becomes larger for larger electron degeneracy.
The higher $\nu_e$ energy is due to the fact that the forward emission of $\nu_e$ against electrons
is favored in the pair neutrino process \cite{Buras03}.
In the temperature range of the Si core burning ($9.5 \lesssim \log T_{\rm C} \lesssim 9.6$)
the average $\bar{\nu}_e$ temperature is less than the threshold energy 1.8 MeV of 
$p (\bar{\nu}_e, e^+)n$ reaction.

\section{Properties of neutrinos emitted from presupernova stars}

\subsection{15 $M_\odot$ model}

The evolution of the central core during the final stage is qualitatively in common among 
12--20 $M_\odot$ stars.
We present neutrino spectra and the structure in five different stages listed in Table \ref{tab:5pt} 
in the 15 $M_\odot$ model.

\begin{figure}
\includegraphics[width=6cm,angle=270]{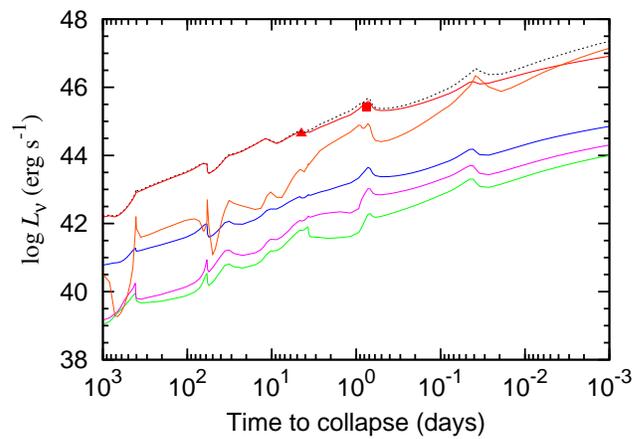}
\caption{
Time evolution of neutrino luminosity until the onset of the core-collapse
($\log T_{\rm C}$ = 9.8) of the 15 $M_\odot$ model.
The dotted line indicates the total luminosity.
Red, blue, pink, green, and orange lines are the contributions of pair neutrinos, photo neutrinos, 
neutrino Bremsstrahlung, plasma neutrinos, and weak interactions of nuclei. 
Triangle and rectangle correspond to the ignition and termination of the Si core burning,
respectively.
}
\label{fig:lnu}
\end{figure}

\subsubsection{Time evolution of neutrino emission}

First, we show the contributions of the neutrino emission processes adopted 
in the 15 $M_\odot$ stellar evolution models to the energy loss by neutrinos.
Figure \ref{fig:lnu} shows the time variation of the neutrino luminosity by the above 
neutrino emission processes from the central Ne burning to the collapse.
Pair neutrino process dominates the neutrino luminosity for most of the advanced stellar evolution.
For last several minutes, the luminosity of weak interaction reactions of nuclei exceeds 
that of pair neutrino process.
Note that pair neutrino process produces all flavors of neutrinos.
However, weak interactions of nuclei mainly produce $\nu_e$ because electron captures rather
than $\beta^-$-decays occur in the collapsing iron core.
Since current neutrino detectors such as KamLAND and Super-Kamiokande mainly detect
$\bar{\nu}_e$ events through $p(\bar{\nu}_e,e^+)n$ reaction, the main sources of the 
neutrino events from preSN stars will be pair neutrinos.
Investigating properties of neutrinos produced through weak interactions of nuclei
is beyond the scope of this study.

\begin{figure}
\includegraphics[width=6cm,angle=270]{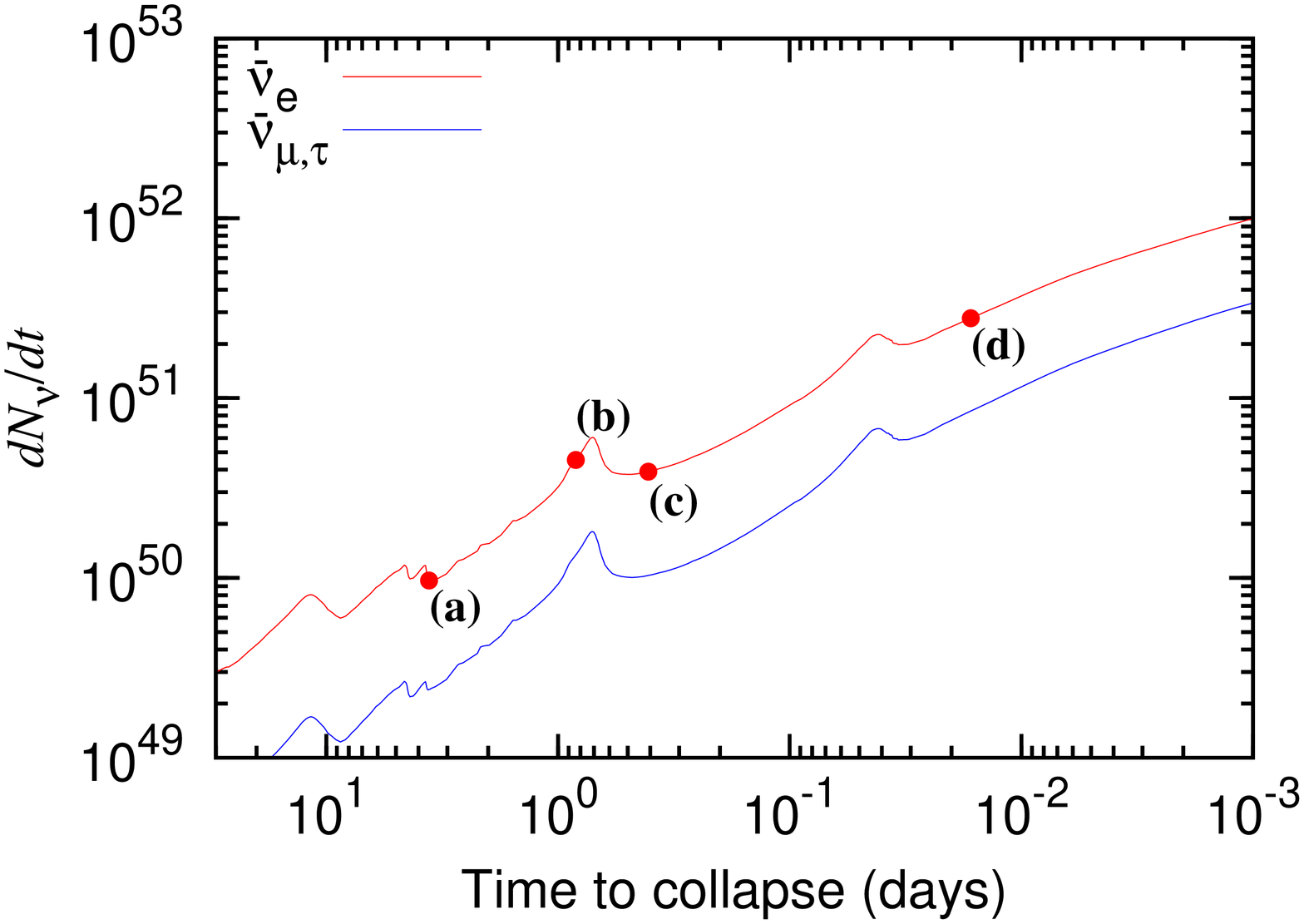}
\includegraphics[width=6cm,angle=270]{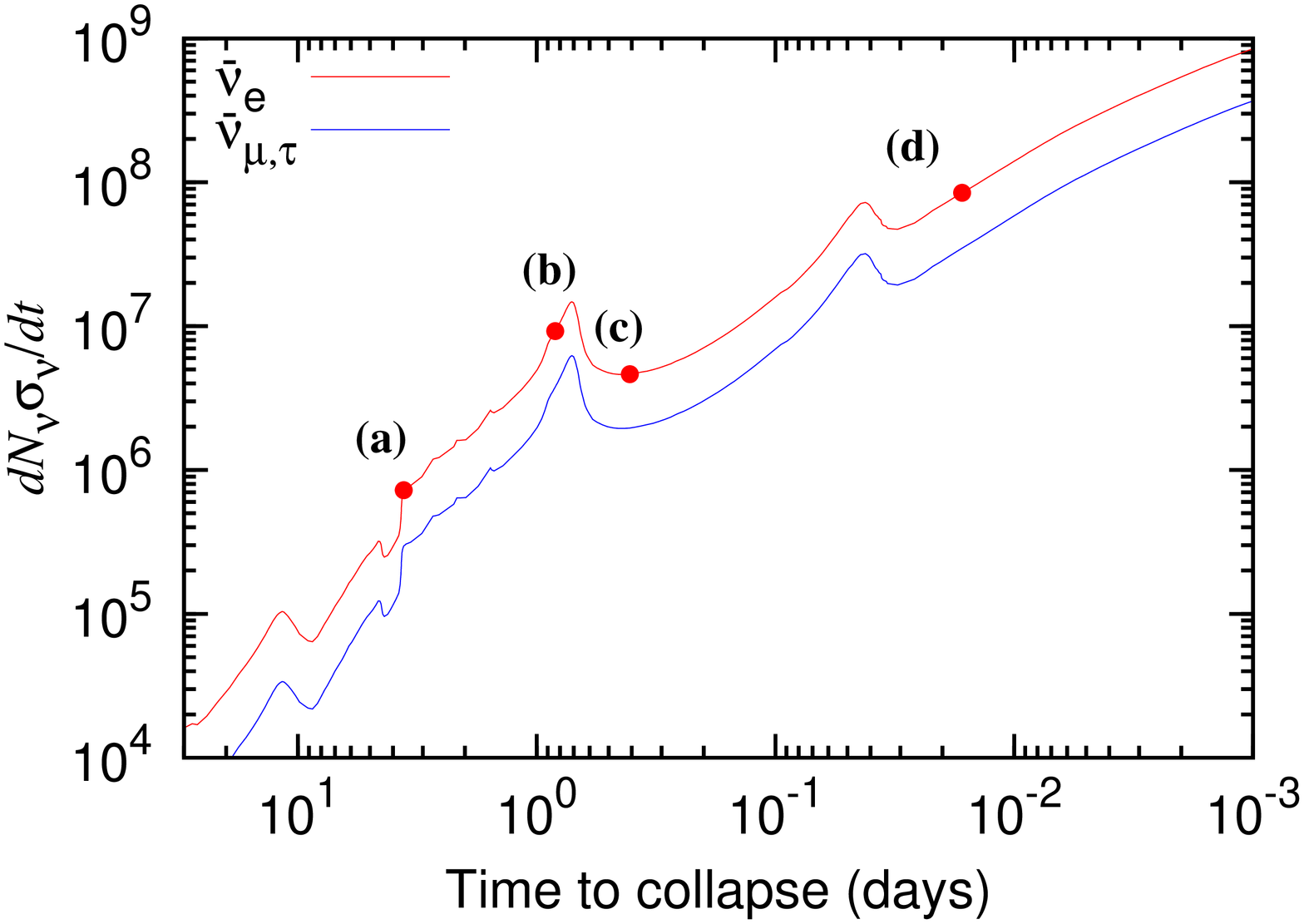}
\caption{
Time evolution of the emission rates for $\bar{\nu}_e$ and $\bar{\nu}_{\mu,\tau}$ 
in the 15 $M_\odot$ model.
Top panel shows the neutrino emission rates.
Bottom panel shows the rate $dN_{\bar{\nu}_{\alpha}} \sigma_{\nu}/dt$ defined in Equation (\ref{eq:rate}).
Points (a)--(d) indicate different evolution stages explained in Sec. III A 2.
\label{fig:dNnudt}
}
\end{figure}

Next, we show the time evolution of the pair neutrino emission rate.
The top panel of Fig. \ref{fig:dNnudt} shows the time evolution of the emission rate for  $\bar{\nu}_e$ and 
$\bar{\nu}_{\mu,\tau}$ produced through pair neutrino process.
Note that the neutrino emission rate of $\bar{\nu}_{\mu,\tau}$ is the sum of the rates of $\bar{\nu}_{\mu}$
and $\bar{\nu}_{\tau}$.
The neutrino emission rates increase with time for most of the time because the star gradually
contracts and the temperature in the central region rises.
On the other hand, the rates decrease temporally
when the main burning process changes.
The Si core burning ignites before point (a).
The O shell burning starts between points (b) and (c).
The Si shell burning around $M_r \sim 1 M_\odot$ starts at a time just before point (d).
At the ignitions of the Si core burning, the O shell burning, and the Si shell burning, 
the $\bar{\nu}_e$ emission rate decreases by factors of 1.2, 1.6, and 1.1, respectively.

The neutrino event rate is determined by the multiple of the neutrino emission rate and the 
neutrino cross section.
Most current and future neutrino detectors detect $\bar{\nu}_e$ from preSN stars
through the $p(\bar{\nu}_e, e^+)n$ reaction.
Therefore, it is useful to evaluate the quantity considering the weight of the neutrino
cross section to the neutrino emission rate.
We consider the rate $dN_{\bar{\nu}_{\alpha}} \sigma_{\nu}/dt$ of $\bar{\nu}_\alpha$ defined by
%\begin{equation}
%\frac{dN_{\bar{\nu}_{\alpha}}\sigma_\nu}{dt} =
%\int_0^\infty \frac{dN_{\bar{\nu}_{\alpha}}(\varepsilon_\nu,t)}{d\varepsilon_\nu dt} \sigma_{p+\bar{\nu}_e}
%(\varepsilon_\nu) d\varepsilon_{\nu},
%\label{eq:rate}
%\end{equation}
\begin{eqnarray}
\frac{dN_{\bar{\nu}_{\alpha}}\sigma_\nu}{dt} &=&
\int_0^{M} \frac{d^2N_{\bar{\nu}_{\alpha}}\sigma_\nu(t,M_r)}{dt dM_r} dM_r \\
&=& 
\int_0^{M} \left\{ \int_0^\infty \frac{d^3N_{\bar{\nu}_{\alpha}}(t, M_r, \varepsilon_\nu)}{dt dM_r d\varepsilon_\nu} \sigma_{p+\bar{\nu}_e}
(\varepsilon_\nu) d\varepsilon_{\nu} \right\} dM_r, \nonumber
\label{eq:rate}
\end{eqnarray}
where $d^3N_{\bar{\nu}_{\alpha}}(t, M_r, \varepsilon_\nu)/d\varepsilon_\nu dM_r dt$ is the emission rate of 
$\bar{\nu}_{\alpha}$ with the neutrino energy $\varepsilon_\nu$, at the mass coordinate $M_r$, 
and at a time $t$,  $\sigma_{p + \bar{\nu}_e}(\varepsilon_\nu)$ is the cross section of $p(\bar{\nu}_e,e^+)n$
as a function of the neutrino energy, and $M$ is the mass of the star.
We call $dN_{\bar{\nu}_{\alpha}} \sigma_{\nu}/dt$ ^^ ^^ the detected $\bar{\nu}_{\alpha}$ emission rate"
and call $d^2N_{\bar{\nu}_{\alpha}} \sigma_{\nu}/dM_r dt$ the detected $\bar{\nu}_{\alpha}$ emission 
rate at the mass coordinate $M_r$.
The cross section of $p(\bar{\nu}_e, e^+)n$ is adopted from \cite{Strumia03}.
The threshold energy of this reaction is 1.8 MeV.

The bottom panel of Fig. \ref{fig:dNnudt} shows the time evolution of the detected neutrino emission rate.
This rate also increases toward the collapse for most of time.
This rate steeply rises at the ignition of the Si core burning [see point (a)], 
whereas the emission rate drops.
Before the Si core burning, the neutrinos are mainly emitted from off-centered region 
where the O shell burning proceeds.
When the Si core burning ignites, the main neutrino emission region changes to the center.
Since the temperature in the main neutrino emission region becomes high, 
the average neutrino energy also becomes high.
Combining to larger cross section for higher energy neutrinos, 
the detected emission rate rises steeply by a factor of 1.9 at the Si core ignition.

On the other hand, the drop of the rate $dN_{\bar{\nu}_{\alpha}} \sigma_{\nu}/dt$ 
between points (b) and (c) is more prominent than that of the neutrino emission rate.
The rate drops by a factor of 3.2.
During the stellar evolution in this period, the region of the main neutrino emission
changes from the stellar center to the outer region $M_r \sim 1.4 M_\odot$ 
where the O shell burning occurs.
Since the temperature of the O shell burning region is lower than the central temperature,
the neutrinos from the outer region are more difficult to be detected than the neutrinos from the center.
More details will be explained in Sec. III A 2.

\begin{figure*}
\includegraphics[width=14cm]{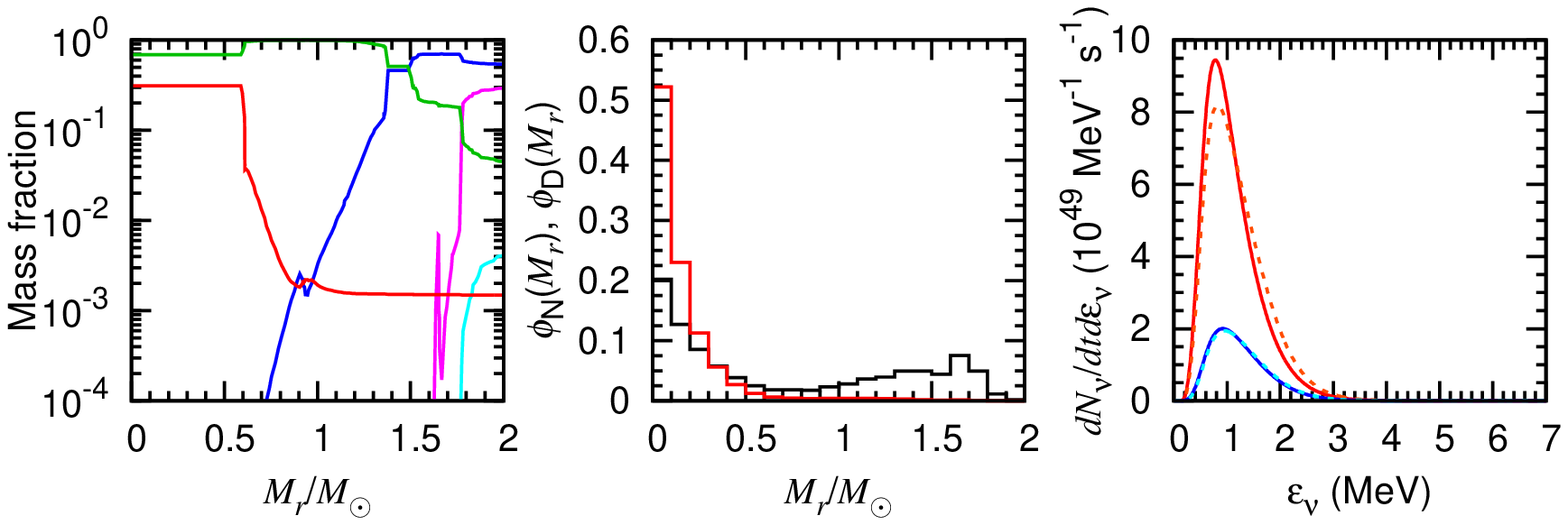}
\caption{
Mass fraction distribution and neutrino properties during the central Si burning 
(3.60 days before the last step; point (a) in Figs. \ref{fig:rhoctc} and \ref{fig:dNnudt}).
Left panel: mass fraction distribution in $M_r \le 2.0 M_\odot$.
Red, green, magenta, blue, and cyan lines indicate ^^ ^^ Fe," ^^ ^^ Si," Ne, O, and C, respectively.
Center panel: the $\bar{\nu}_e$ emission fraction and the detected $\bar{\nu}_e$ fraction.
Black line indicates the $\bar{\nu}_e$ emission fraction, $\phi_{\rm N}(M_r)$, and 
red line indicates the detected $\bar{\nu}_e$ fraction, $\phi_{\rm D}(M_r)$.
See text for details.
Right panel: spectra of $\nu_e$ (orange dashed line), $\nu_{\mu,\tau}$ (cyan dashed line), 
$\bar{\nu}_e$ (red solid line), and $\bar{\nu}_{\mu,\tau}$ (blue solid line).
\label{fig:spc-a}
}
%\end{figure*}
%\begin{figure*}[b]
\includegraphics[width=14cm]{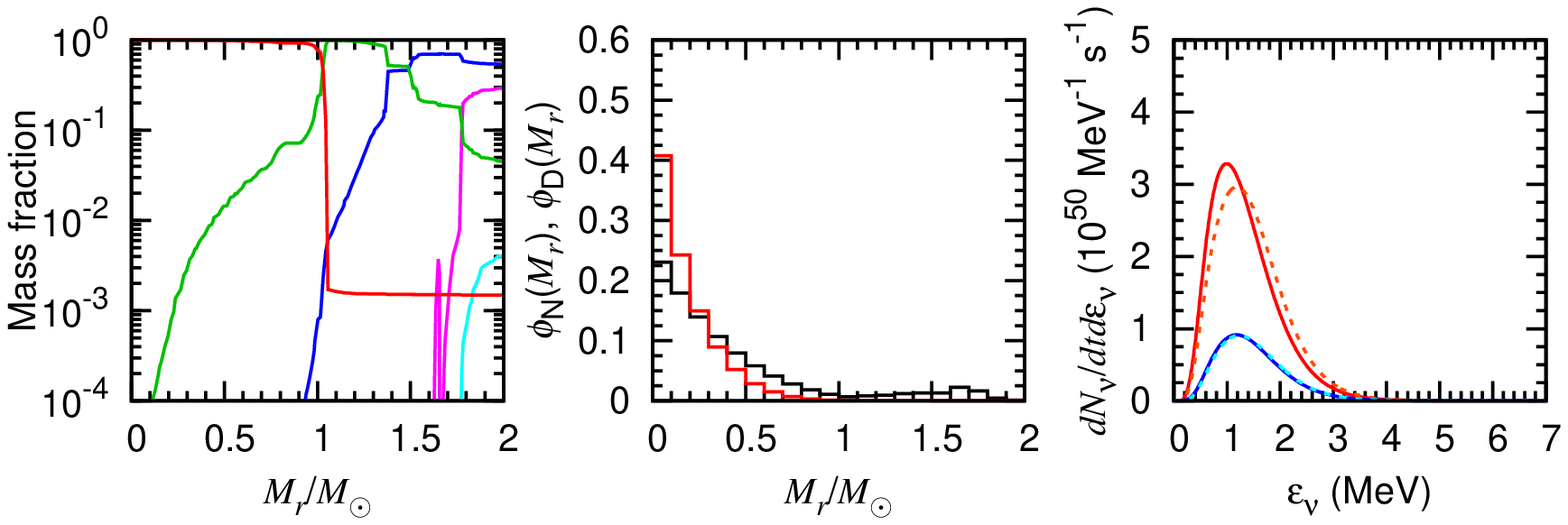}
\caption{
Same as Fig. \ref{fig:spc-a} but for the stage after the central Si burning (20.1 hours before the last step;
point (b) in Figs. \ref{fig:rhoctc} and \ref{fig:dNnudt}).
\label{fig:spc-b}
}
\end{figure*}

\subsubsection{Neutrino energy spectra}

We present neutrino properties at five different stages (a)--(e) in Table \ref{tab:5pt}
during the final evolution stage of the 15 $M_\odot$ model.
The central temperature and density, and the neutrino emission rate in these stages
are shown in Figs. \ref{fig:rhoctc} and \ref{fig:dNnudt}.
Here, we will show the mass fraction distribution, the fraction of the $\bar{\nu}_e$ emission rate
in the region of $M_r \le 2 M_\odot$, and the neutrino spectra.
We define the fractions of the $\bar{\nu}_e$ emission rate $\psi_{\rm N}(M_{\rm in}, M_{\rm out})$ and 
the detected $\bar{\nu}_e$ rate $\psi_{\rm D}(M_{\rm in}, M_{\rm out})$ in the mass range between
$M_{\rm in}$ and $M_{\rm out}$ as
\begin{equation}
\psi_{\rm N}(M_{\rm in}, M_{\rm out}) = 
\frac{\int_{M_{\rm in}}^{M_{\rm out}} \frac{d^2 N_{\bar{\nu}_e}(t, M_r')}{dt dM_r'} dM_r'}
{\int_{0 M_\odot}^{2 M_\odot} \frac{d^2 N_{\bar{\nu}_e}(t, M_r')}{dt dM_r'} dM_r'}
\end{equation}
and
\begin{equation}
\psi_{\rm D}(M_{\rm in}, M_{\rm out}) = 
\frac{\int_{M_{\rm in}}^{M_{\rm out}} \frac{d^2 N_{\bar{\nu}_e}\sigma_\nu(t, M_r')}{dt dM_r'} dM_r'}
{\int_{0 M_\odot}^{2 M_\odot} \frac{d^2 N_{\bar{\nu}_e}\sigma_\nu(t, M_r')}{dt dM_r'} dM_r'} ,
\end{equation}
where $d^2 N_{\bar{\nu}_e}(t, M_r)/dt dM_r$ is the $\bar{\nu}_e$
emission rate at the mass coordinate $M_r$, and $d^2 N_{\bar{\nu}_e}\sigma_\nu(t, M_r)/dt dM_r$ 
is the detected $\bar{\nu}_e$ rate defined at Eq. (3).
We also define the $\bar{\nu}_e$ emission fraction $\phi_{\rm N}(M_r)$ and 
the detected $\bar{\nu}_e$ fraction $\phi_{\rm D}(M_r)$ in the interval of 0.1 $M_\odot$ below the 
mass coordinate $M_r$ as
\begin{equation}
\phi_{({\rm N, D})}(M_r) = \psi_{({\rm N, D})}(M_r - 0.1 M_\odot, M_r).
\end{equation}
They are indicators of the location where electron antineutrinos are mainly emitted from.
In the following figures on the fractions of the $\bar{\nu}_e$ emission rate and 
the detected $\bar{\nu}_e$ rate, we will show the distributions of $\phi_{\rm N}(M_r)$ and 
$\phi_{\rm D}(M_r)$, respectively, 
taking discrete values with the interval of 0.1 $M_\odot$ for $M_r$, 
i.e., $M_r = 0.1$, 0.2, ..., 2.0 $M_\odot$.

Stage (a) is at 3.61 days before the last step.
This star is in the convective Si core burning.
The left panel of Fig. \ref{fig:spc-a} shows the mass fraction distribution in $M_r \le 2 M_\odot$.
The convection region extends to 0.57 $M_\odot$.
The center panel shows the $\bar{\nu}_e$ emission fraction and the detected $\bar{\nu}_e$ fraction.
There is a peak at the center for the both fractions.
Electron antineutrinos of 54 \% are emitted from the convective Si/Fe core. 
The rest is emitted from the surrounding Si and O-rich layers.
On the other hand, almost all electron antineutrinos detected through $p(\bar{\nu}_e,e^+)n$ are 
emitted from the convective Si/Fe core.
About half are emitted from the central region, $\phi_{\rm D}(0.1M_\odot) = 0.52$.
The right panel shows the neutrino energy spectra.
The neutrino emission rate is the order of $10^{50}$ s$^{-1}$ at this time.
The average energy is 1.1--1.3 MeV, that is smaller than the threshold energy of 
$p(\bar{\nu}_e,e^+)n$.

Stage (b) is at 20.1 hours before the last step.
The Si core burning has ceased 1.2 hours before this time and the central region of 
the star is contracting.
The left panel of Fig. \ref{fig:spc-b} indicates that the Fe core of 1.02 $M_\odot$ has been formed.
The intermediate elements denoting ^^ ^^ Si" have been exhausted through the Si core burning.
The surrounding layer enriched in ^^ ^^ Si" extends to 1.37 $M_\odot$.
The O mass fraction increases outwards there.
In the center panel, most of electron antineutrinos are emitted from the Fe core and the detected ones 
are from the central region of the core.
We obtain $\psi_{\rm N}(0M_\odot, 1.0M_\odot) = 0.89$ and 
$\psi_{\rm D}(0M_\odot, 0.5M_\odot) = 0.94$.
The right panel indicates that the neutrino emission rate increases by a factor of four
from the previous stage.
The average energy slightly increases from the Si core burning but is still
smaller than the threshold energy of $p(\bar{\nu}_e,e^+)n$.

\begin{figure*}
\includegraphics[width=14cm]{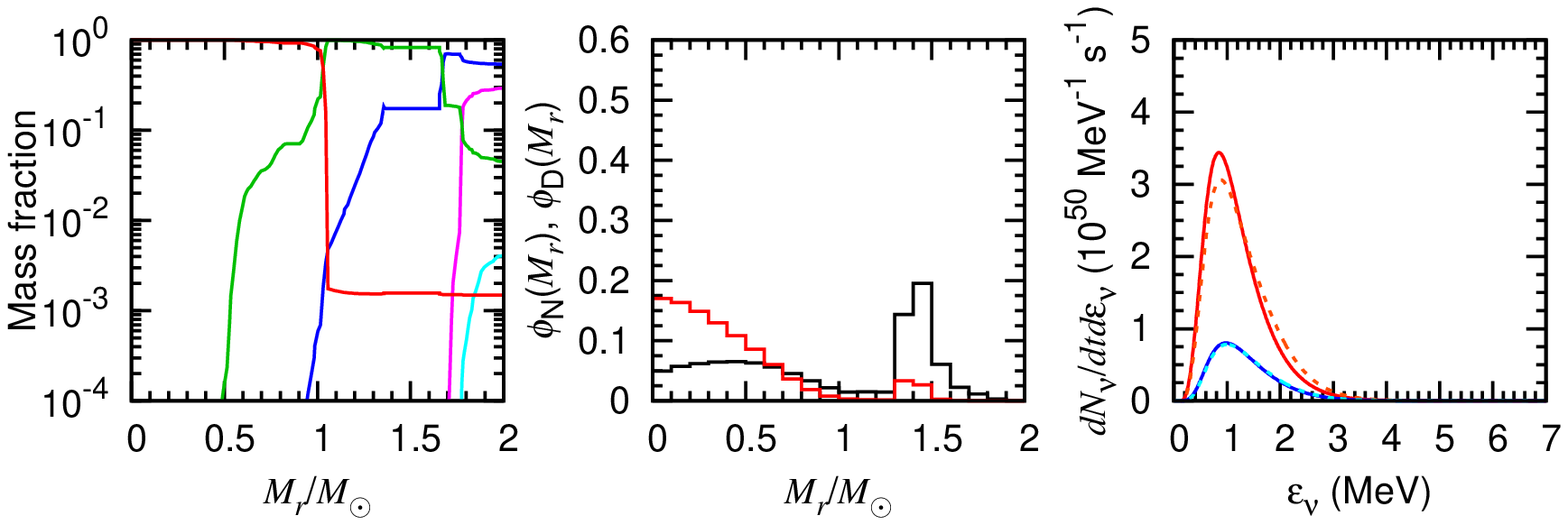}
\caption{
Same as Fig. \ref{fig:spc-a} but for the stage at 9.78 hours before the last step 
[point (c) in Figs. \ref{fig:rhoctc} and \ref{fig:dNnudt}].
\label{fig:spc-c}
}
%\end{figure*}
%\begin{figure*}
\includegraphics[width=14cm]{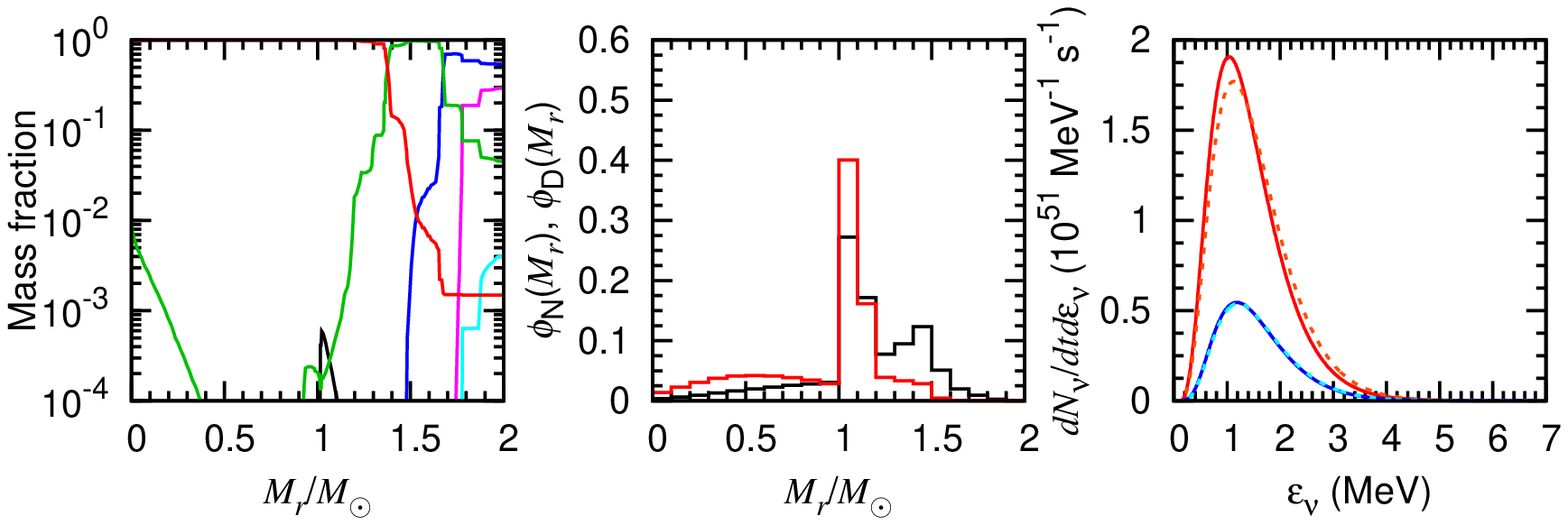}
\caption{
Same as Fig. \ref{fig:spc-a} but for the stage at 23.8 minutes before the last step 
[point (d) in Figs. \ref{fig:rhoctc} and \ref{fig:dNnudt}]
In the left panel, black line indicates He.
\label{fig:spc-d}
}
\end{figure*}

Stage (c) is at 9.78 hours before the last step.
Before this time Oxygen ignited at $M_r \sim 1.36 M_\odot$ and the convective Si/O layer extended
outwards.
The O shell burning expands the star and the central temperature and density decrease
[see the range between (b) and (c) in Fig. \ref{fig:rhoctc}].
The neutrino emission rate also decreases at that time (see Fig. \ref{fig:dNnudt}).
We see in the left panel of Fig. \ref{fig:spc-c} that the mass fractions of Si and O are almost constant 
between 1.36 and 1.66 $M_\odot$.
The convective O shell burning occurs in this region.
The temperature in this region rises and the high temperature
raises the neutrino emissivity.

The center panel indicates that the $\bar{\nu}_e$ emission is not concentrated to the center and it
is broadly distributed.
We also see a peak in the O shell burning region for $\phi_{\rm N}(M_r)$.
We see the decrease in $\psi_{\rm N}(0M_\odot, 1.0M_\odot)$ to 0.52 and 
the increase in $\psi_{\rm N}(1.3M_\odot, 1.6M_\odot)$ to 0.41.
However, the distribution of the detected $\bar{\nu}_e$ fraction has a peak at the center and 
the contribution of the convective Si/O layer is still small.
Most of the detected $\bar{\nu}_e$ are from the Fe core, i.e., 
$\psi_{\rm D}(0M_\odot, 1.0M_\odot) = 0.93$ and $\psi_{\rm D}(1.3M_\odot, 1.6M_\odot) = 0.06$.
This is because the electron antineutrinos emitted from the central region have higher energy than
the ones from the Si/O layer.

The right panel shows the neutrino spectra at this time.
Comparing with the spectra in the previous stage, we do not see clear difference
 in the maximum value for each neutrino flavor.
 On the other hand, the spectra become less energetic from the previous stage.
This is due to the contribution of the neutrinos from the Si/O layer.
Thus, when the O shell burning starts, the neutrino 
energy spectra become less energetic and the detectability of the neutrino events
will decrease.

Stage (d) is at 23.8 minutes before the last step.
After stage (c), the O shell burning ceases, the star contracts again, and 
the Si shell burning ignites at $M_r \sim 1 M_\odot$.
The Fe core and the surrounding Si layer grow up through the shell burnings.
The left panel of Fig. \ref{fig:spc-d} shows that the Fe core mass and the outer boundary of 
the Si layer become 1.37 and 1.65 $M_\odot$, respectively.
We also see the O/Si-rich and O/Ne-rich layers outside the Si layer.

In the center panel, we see two peaks at $M_r \sim 1$ and 1.4 $M_\odot$ for the $\bar{\nu}_e$ emission.
At this stage, the central temperature is $\log T_{\rm C} = 9.66$ and 
there are two peaks at $M_r = 1.02$ and 1.36 $M_\odot$ in the temperature distribution.
At $M_r = 1.02 M_\odot$ the temperature is $\log T = 9.64$ and the density is $\log \rho = 6.79$.
The high temperature and low density in this region produce the large neutrino emission rate.
The $\bar{\nu}_e$ emission rate in this region is much larger than that at the center
(see Fig. \ref{fig:qnucnt}).
Different from the stage (c), we also see a peak for the detected $\bar{\nu}_e$ fraction in this region.
This corresponds to $\phi_{\rm D}(1.0M_\odot, 1.2M_\odot) = 0.56$.
Some detected electron antineutrinos are also produced inside the Si shell; 
$\phi_{\rm D}(0M_\odot, 1.0M_\odot) = 0.33$.

The right panel shows that the neutrino emission rate increases to the order of $10^{51}$ s$^{-1}$.
Although the neutrino emission rate decreases when the Si shell burning starts,
the star has turned to the contraction at this stage.
The neutrino spectra shift to higher energy from the previous stage.
This is also due to the contraction after the O and Si shell burnings.

\begin{figure*}
\includegraphics[width=14cm]{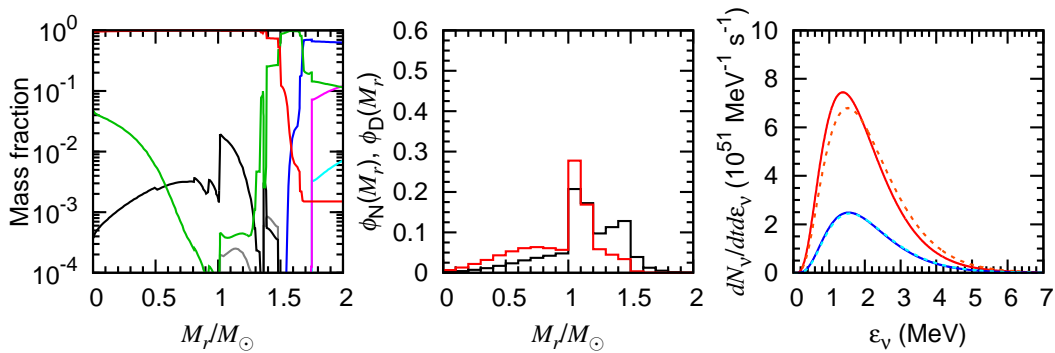}
\caption{
Same as Fig. \ref{fig:spc-a} but for the last step 
(point (e) in Figs. \ref{fig:rhoctc} and \ref{fig:dNnudt}).
In the left panel, black and grey lines indicate He and H.
\label{fig:spc-e}
}
\end{figure*}

Stage (e) is the last step of the stellar evolution calculation.
The central region including the Fe core and Si and O/Si shells is contracting and 
the central temperature exceeds $\log T_{\rm C} = 9.81$.
The left panel of Fig. \ref{fig:spc-e} shows the mass fraction distribution at this time.
The Fe core grows to 1.50 $M_\odot$ through the Si shell burning.
The distributions of the $\bar{\nu}_e$ emission fraction and the detected $\bar{\nu}_e$ fraction, 
seen in the center panel,  are similar to stage (d);
there are two peaks at $M_r \sim 1.0$ and 1.4 $M_\odot$ for the $\bar{\nu}_e$ emission
and a peak at $M_r \sim 1.0 M_\odot$ for the detected $\bar{\nu}_e$.
The two peaks of the temperature distribution seen in the previous stage still remain
and the peak temperatures rise.
The detected rate of $\bar{\nu}_e$ from the Si shell burning region is slightly larger than 
the rate of $\bar{\nu}_e$ from inside the region.
We obtain $\phi_{\rm D}(0M_\odot, 1.0M_\odot) = 0.41$ and 
$\phi_{\rm D}(1.0M_\odot, 1.2M_\odot) = 0.45$ at this stage.
The right panel shows that the emission rates of $\nu_e$ and $\bar{\nu}_e$ increase to 
about $10^{52}$ s$^{-1}$ and the spectra shift to high energy direction further.
The central high density region contributes the production of high energy neutrinos.
The average energy of the neutrino spectra exceeds the threshold energy of $p(\bar{\nu}_e,e^+)n$
reaction (see Table \ref{tab:5pt}).

\subsection{Stellar mass dependence of neutrino emission}

Massive stars form an Fe core in their final stage of the evolution and the evolution
time scale depends on the initial mass.
The period of the Si core burning is 7.4 days and 16 hours for the 12 and 20 $M_\odot$ models,
respectively (see Table \ref{tab:stars}).
Thus, quantitative properties of neutrinos emitted in the final stage also depend on the stellar mass.
Here, we show the dependence of the neutrino emission rate and average energy on the stellar mass.

\begin{figure}[b]
\includegraphics[width=6cm,angle=270]{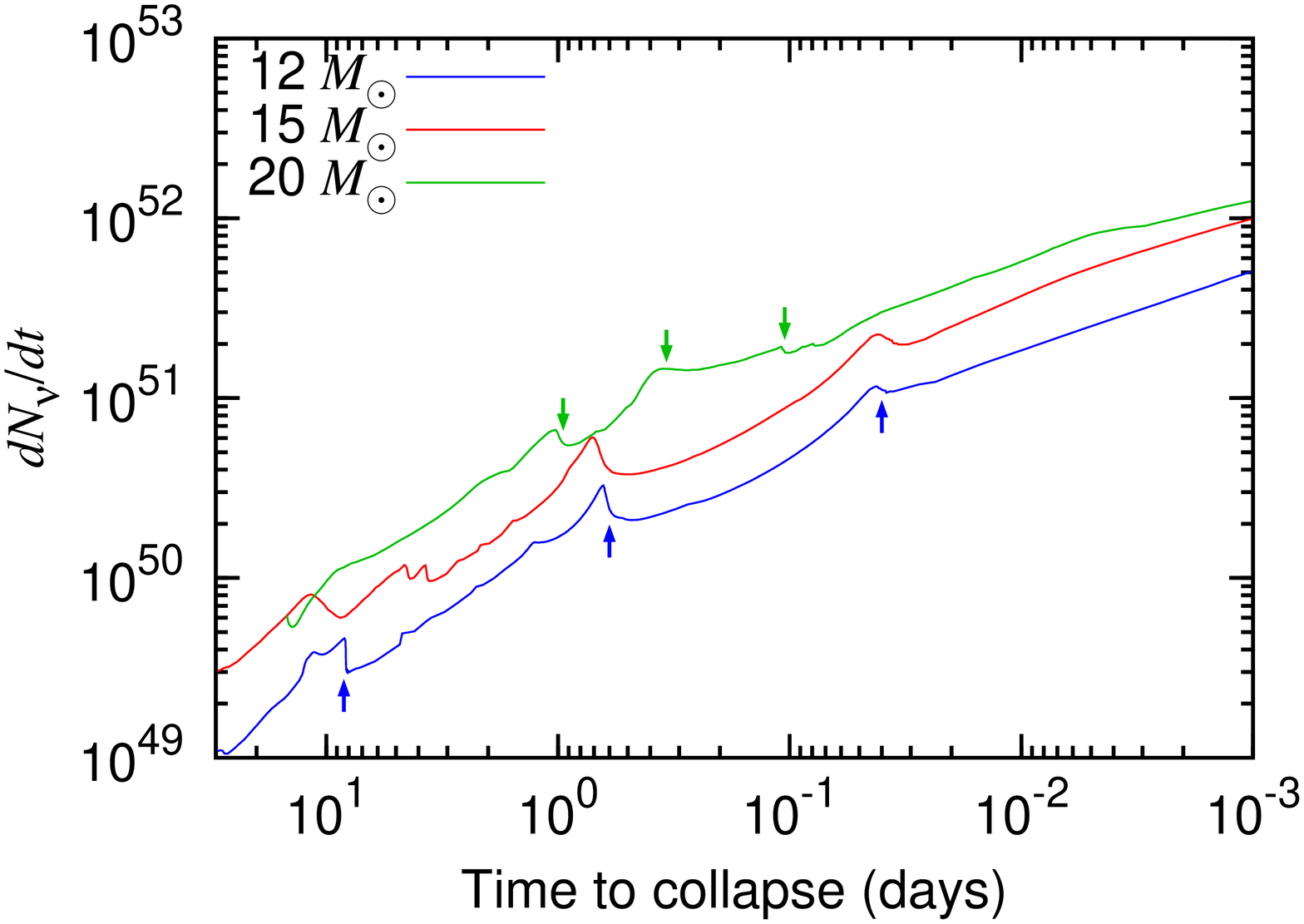}
\includegraphics[width=6cm,angle=270]{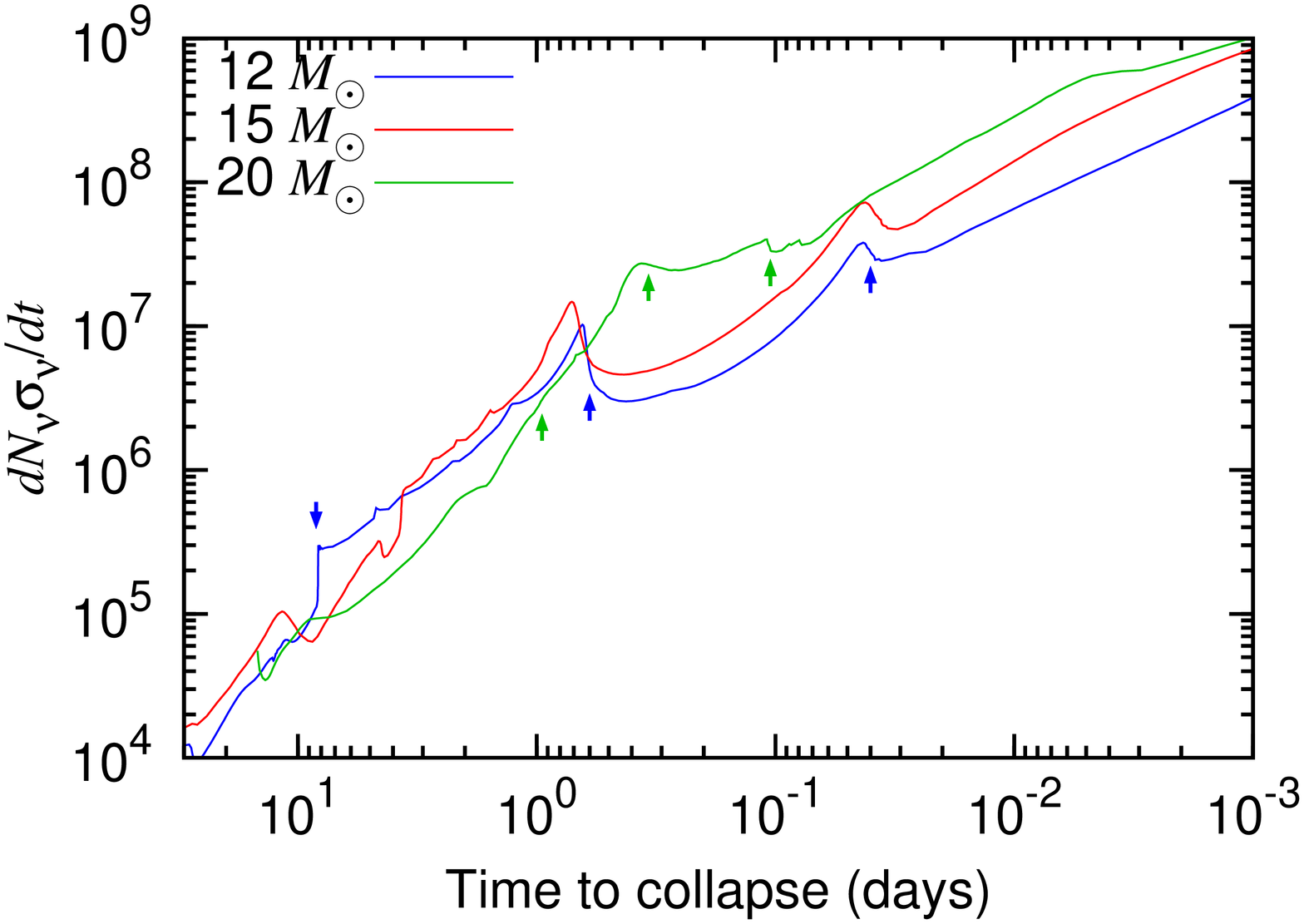}
\caption{
Time evolution of the $\bar{\nu}_e$ emission rate (top panel) and the detected $\bar{\nu}_e$
emission rate (bottom panel) 
of the 12 (blue line), 15 (red line), and 20 (green line) $M_\odot$ models.
Blue and green arrows indicate the time when the neutrino emission rate decreases in the 12 and 
20 M$_\odot$ models, respectively.
See text for details.
\label{fig:dNnudt3}
}
\end{figure}

Figure \ref{fig:dNnudt3} shows the time evolution of the $\bar{\nu}_e$
emission rate and the detected $\bar{\nu}_e$ emission rate of the 12, 15, and 20 $M_\odot$ model.
Since the stellar mass dependence of the emission rate for other flavors is similar to that
of $\bar{\nu}_e$, we discuss only the dependence of  $\bar{\nu}_e$.
Both the $\bar{\nu}_e$ emission rate and the detected $\bar{\nu}_e$ emission rate increase
with time for most of the evolution period among the three models.
In the 12 $M_\odot$ model, the $\bar{\nu}_e$ emission rate is smaller than the 15 $M_\odot$ model
for a given time.
The $\bar{\nu}_e$ emission rate decreases by a factor of 1.5 (see the left blue arrow in the top panel) 
and the detected $\bar{\nu}_e$ rate increases by a factor of 2.5 (see the left blue arrow in the bottom panel) 
around 8.2 days before the last step.
At this time the Si core burning starts.
Although this change is also seen in the 15 $M_\odot$ model, the corresponding time is different.
This is due to the longer period of the Si core burning in the 12 $M_\odot$ model.

We see the decrease in both rates in 10 hours before the last step, assigned by the center blue arrow 
in each panel.
The decreases in the $\bar{\nu}_e$ emission rate and the detected emission rate are 
factors of 1.6 and 3.4, respectively.
This is evidence for the ignition of the O shell burning.
In 50 minutes before the last step, we also see slight decreases in both rates, assigned by 
the right blue arrow.
The $\bar{\nu}_e$ emission rate and the detected emission rate decrease by factors of 1.1 and 1.3, 
respectively.
This corresponds to the ignition of the Si shell burning.
The trend of these decreases is similar to the 15 $M_\odot$ model.

In the 20 $M_\odot$ model, the $\bar{\nu}_e$ emission rate is generally larger than 
the 15 $M_\odot$ model and the neutrino emission rate has some properties different from
the other stellar models.
We see the decrease in the emission rate is seen by a factor of 1.2 at 0.9 days before the last step
(the left green arrow in the top panel).
However, the steep increase in the detected $\bar{\nu}_e$ emission rate is not seen at that time
(the left green arrow in the bottom panel).
The $\bar{\nu}_e$ emission rate slightly decreases at 6.5 and 2.4 hours before the last step
(see the center and right arrows in each panel).
The corresponding decreases of the detected $\bar{\nu}_e$ emission rate are small.
This seems to be due to differences of burning processes in the advanced evolution.
In the 20 $M_\odot$ model, the convective shell of the O shell burning before the Si core burning 
extended to 1.5 $M_\odot$ and the oxygen in this region has been exhausted.
The Si shell grows up to 1.59 $M_\odot$ when the Si core burning ended.
Then, the O shell burning occurs in the region of  $M_r \gtrsim 1.6 M_\odot$.
However, this burning scarcely prevents the core contraction and, thus, the neutrino emission rate 
scarcely decreases.
We do not see the decrease in the central temperature at this time.
We note that, in the 12 and 15 $M_\odot$ models, the O and Si-enriched region remains 
outside $M_r \sim 1.36 M_\odot$ even after the Si core burning and, then, 
the O shell burning in this region is stronger.
The decrease at 2.4 hours before the last step corresponds to the Si shell burning.
The Si shell burning occurs at $M_r \sim 0.7 M_\odot$ that is deeper than the 12 and 15 $M_\odot$
models.
We consider that stronger Si shell burning prevents decreasing the neutrino emissivity.

\begin{figure}
\includegraphics[width=6cm,angle=270]{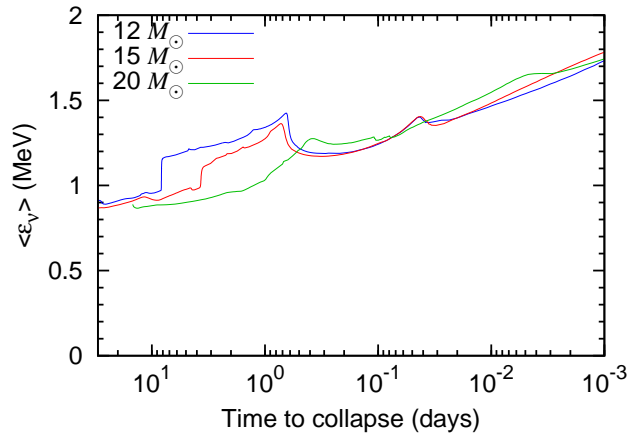}
\caption{
Time evolution of the average $\bar{\nu}_e$ energy 
of the 12 (blue line), 15 (red line), and 20 (green line) $M_\odot$ models.
\label{fig:enuav3}
}
\end{figure}

We show the time evolution of the average $\bar{\nu}_e$ energy 
$\langle \epsilon_{\nu} \rangle$
of these three models in Fig. \ref{fig:enuav3}.
Before the Si core burning, all models indicate the average energy with less than $\sim$ 1 MeV.
When the Si core burning ignites, the neutrino emission from the center increases and 
the average energy becomes large for the 12 and 15 $M_\odot$ models.
The average energy of the 20 $M_\odot$ model also increases more steeply than before,
although the steepness is less than the less massive models.
The average energy during the Si core burning is lower for more massive star models.
As shown in Fig. \ref{fig:rhoctc}, the evolution track of the more massive one passes through 
a lower density and higher temperature path.
The average $\bar{\nu}_e$ energy is higher in higher density for a given temperature.
Thus, we consider that lower density structure of more massive star provides lower average
$\bar{\nu}_e$ energy during the Si burning.
The average energy becomes small during the O shell burning.
The main region of the neutrino emission is the O burning shell, where the temperature
and density is lower than the center.
During the Si shell burning to collapse, we do not see clear dependence on the stellar mass.
This is partly due to the difference of the contraction time scale in these models.
For a given central temperature after the ignition of the Si shell burning, 
the average energy is smaller for more a massive model.

We showed that the time evolution of the neutrino emission rate and the detected neutrino
emission rate from the Si core burning to the core collapse is influenced by burning processes
in the central region of massive stars.
The ignition of the Si core burning raises the average $\bar{\nu}_e$ energy and 
the neutrino detectability.
The onset of O and Si shell burnings decreases the neutrino emission rate as well as the
neutrino detectability for a moment.
The stellar mass dependence of the above burning processes brings about
the dependence of the neutrino properties from preSN stars.

\section{Evaluation of neutrino events by neutrino observatories}

\begin{figure*}
\includegraphics[width=6cm,angle=270]{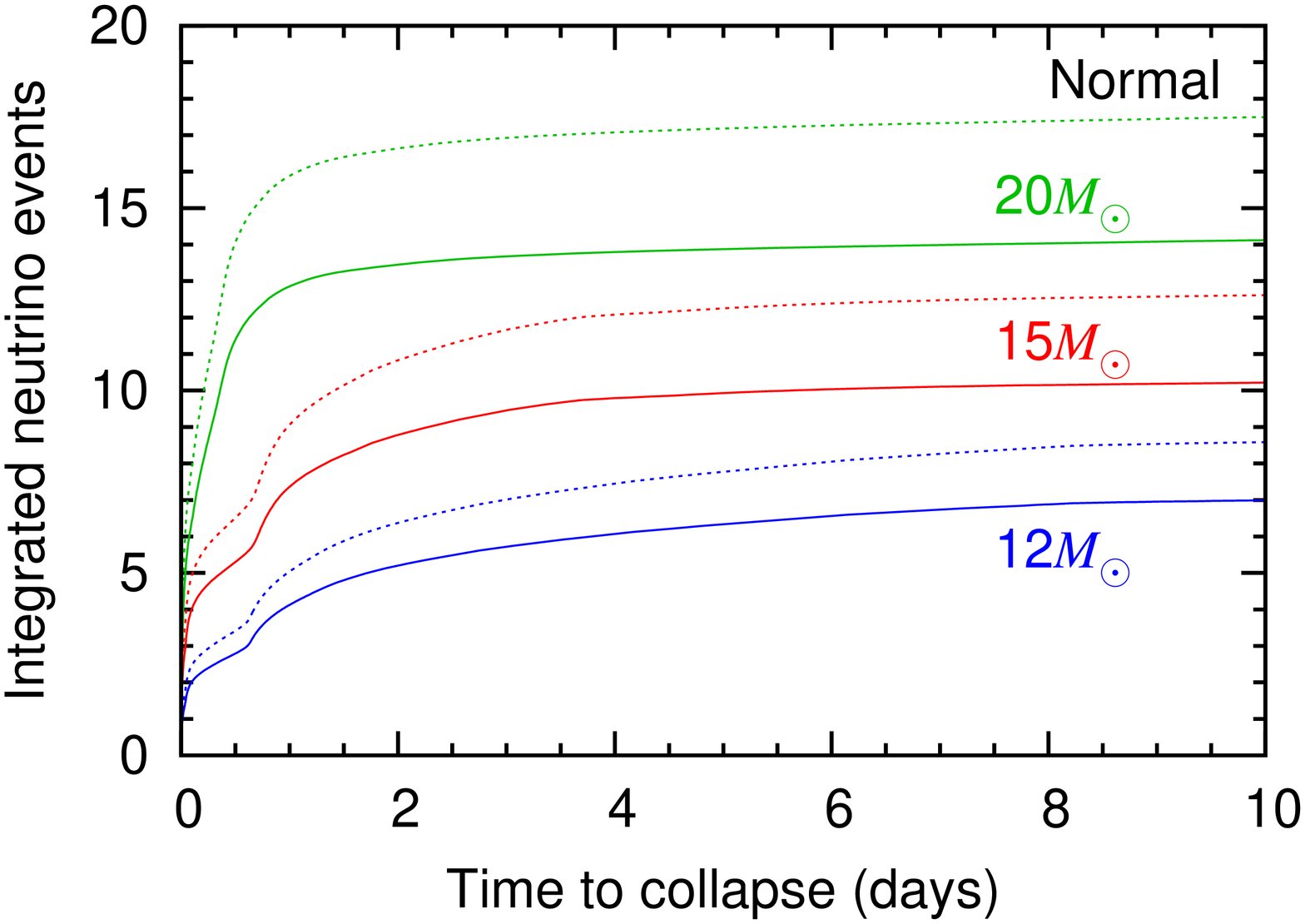}
\includegraphics[width=6cm,angle=270]{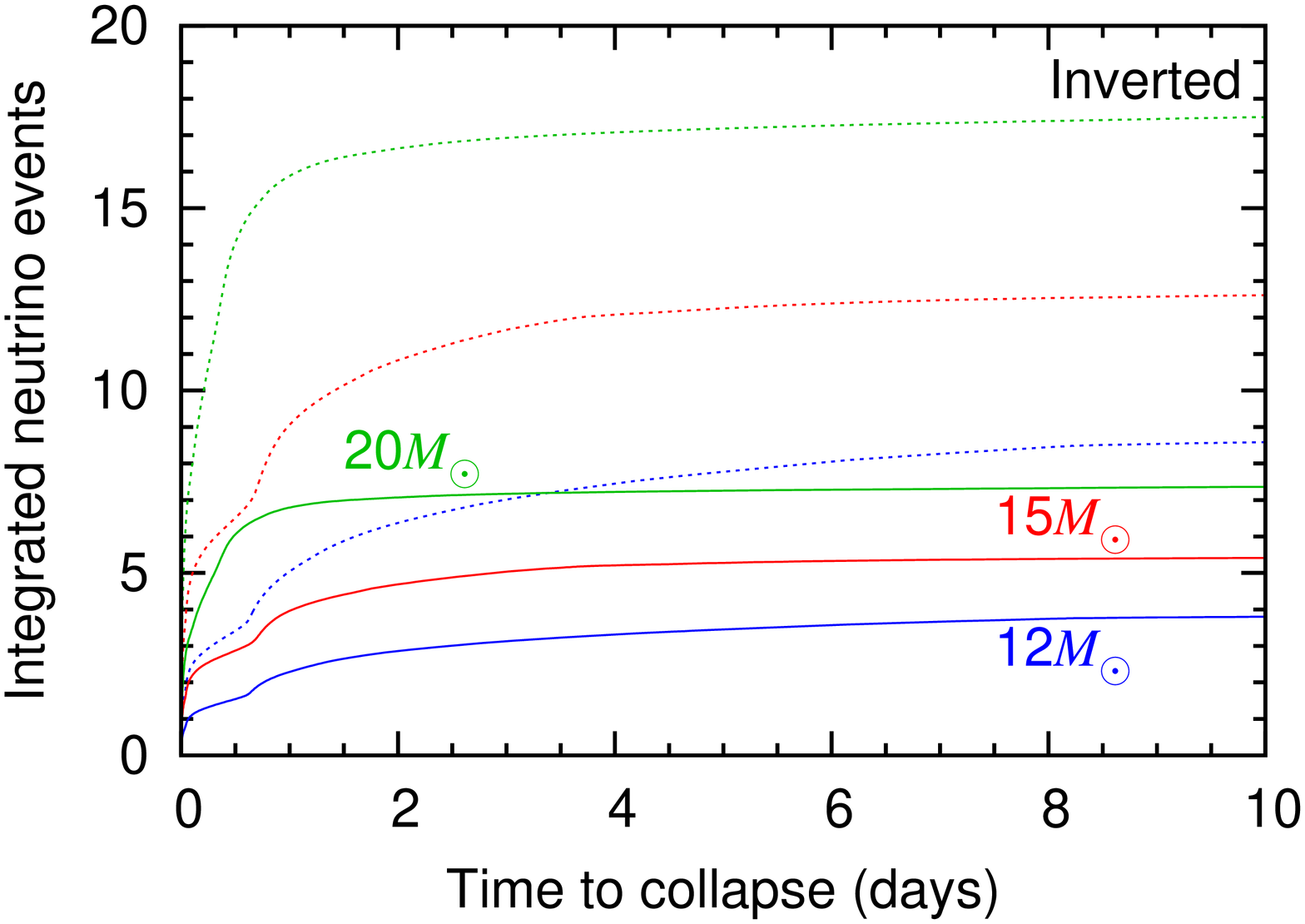}
\caption{
Integrated neutrino events detected by KamLAND.
Left panel corresponds to the normal mass hierarchy and right panel does the inverted
mass hierarchy.
Blue, red, and green lines correspond to the 12, 15, and 20 $M_\odot$ models, respectively.
Dashed lines indicate the cases when the neutrino oscillation is not taken into account.
\label{fig:eventKL}
}
\end{figure*}

We investigate the neutrino event rate and the total events by current and future neutrino detectors.
Current and most future neutrino detectors mainly detect $\bar{\nu}_e$ signals through 
$p(\bar{\nu}_e, e^{+})n$ reaction.
So, we evaluate the neutrino events through this reaction.
In this study, we assume that the distance of a preSN star is 200 pc.
This distance corresponds to the distance to Betelgeuse.

When we investigate the neutrino events of preSN stars, we need to consider the flavor
change by the Mikheyev-Smirnov-Wolfenstein (MSW) effect during the passage of the stellar interior.
The flavor change mainly occurs at resonance layers.
In preSN stars, the flavors change at the high (H) and low (L) resonances 
\cite[e.g.,][]{Dighe00,Takahashi01}.
The adiabaticity and, thus, the transition probability depend on the mixing angles. 
Recent neutrino experiments confirmed that the mixing angle $\sin^2\theta_{13} \sim 0.02$
\cite{AbeY11,AbeK11,Adamson11,An12} and 
the large $\sin^2\theta_{13}$ value indicates adiabatic flavor change at the both resonances.
The density of the H resonance is written as
\begin{equation}
\rho_{\rm H-res} \sim
3.0 \times 10^{4} \left( \frac{\rm 1 MeV}{\varepsilon_\nu} \right) \,  {\rm g \, cm^{-3}}.
\end{equation}
Here, we set squared mass difference between mass eigenstates 1 and 3 as 
$|\Delta m_{31}^2| c^4 = 2.43 \times 10^{-3}$ eV$^2$,  the mixing angle $\theta_{13}$ as 
$\sin^2  \theta_{13} = 0.024$ \cite{Olive14}, and the electron fraction $Y_e$ as 0.5.
Since the density inside the O/Ne layer is larger than the density of the H resonance with
$\varepsilon_\nu > 1$ MeV, it is reasonable to assume that the transition probability does not 
depend on the neutrino energy.
Thus, we evaluate the transition probability of $\bar{\nu}_e \rightarrow \bar{\nu}_e$ 
in normal and inverted mass hierarchies as
\begin{equation}
P_{(\bar{\nu}_{e} \rightarrow \bar{\nu}_{e})} = \left\{ 
 \begin{array}{ll}
\cos^2\theta_{12} \cos^2\theta_{13} = 0.675 & ({\rm normal}) \\
\sin^2\theta_{13} = 0.024 & ({\rm inverted}), 
 \end{array} 
 \right.
\end{equation}
where $\sin^2\theta_{12} = 0.308$ \cite{Olive14}.
The sum of the transition probabilities from $\bar{\nu}_\mu$ and $\bar{\nu}_\tau$ is 
\begin{equation}
\sum_{\alpha=\mu,\tau} P_{(\bar{\nu}_\alpha \rightarrow \bar{\nu}_e)} = 
1 - P_{(\bar{\nu}_e \rightarrow \bar{\nu}_e)}.
\end{equation}

\subsection{KamLAND}

The Kamioka Liquid-scintillator Antineutrino Detector (KamLAND) is a one kton size
neutrino detector located in the Kamioka Mine, Japan.
This detector has a detectability of low energy $\gamma$-rays using its liquid-scintillator.
KamLAND detects $\bar{\nu}_e$ events through $p(\bar{\nu}_e, e^+)n$ from a preSN star.
First, the produced positron is pair-annihilated to produce $\gamma$-rays.
Then, the neutron produced through the neutrino reaction is captured by
a proton through $n(p, \gamma)d$ and 2.2 MeV $\gamma$-rays are emitted.
KamLAND identifies a $\bar{\nu}_e$ event using both the prompt $\gamma$-rays produced by 
the pair-annihilation and the delayed 2.2 MeV $\gamma$-rays produced by the neutron capture.
The low threshold energy enables to detect $\bar{\nu}_e$ events from preSN stars
with the distance of Betelgeuse (e.g., \cite{Kato15,Asakura16}).

We calculate the spectrum of the $\bar{\nu}_e$ events detected by KamLAND in accordance with
\cite{Asakura16}:
\begin{eqnarray}
\frac{d^2 N(t,\varepsilon_{p})}{dt d\varepsilon_{p}} &=& 
\epsilon_{\rm live} \epsilon_{\rm s}(\varepsilon_{p}) \frac{N_{\rm T}}{4 \pi d^2} \\
&\times& \sum_{\alpha} \int 
\frac{d^2N_{\bar{\nu}_\alpha}(t,\varepsilon_{\nu})}{dt d\varepsilon_{\nu}}
P_{(\bar{\nu}_{\alpha} \rightarrow \bar{\nu}_e)} 
\sigma_{p + \bar{\nu}_e}(\varepsilon_{\nu})  \nonumber \\
&\times& \left( \frac{d\varepsilon_{\nu}}{d\varepsilon'_{p}} \right)
R(\varepsilon_{p},\varepsilon'_{p})
d\varepsilon'_{p} , \nonumber
\end{eqnarray}
where $\epsilon_{\rm live}$ is mean livetime-to-runtime ratio, 
$\epsilon_{\rm s}(\varepsilon_p)$ is the total detection efficiency, $N_{\rm T} = 5.98 \times 10^{31}$ 
is the fiducial proton number of KamLAND, $d$ is the distance to a preSN star, that is assumed
to be 200 pc, 
$\varepsilon'_p$ is the expected energy of the prompt event with the relation of 
$\varepsilon'_p = \varepsilon_\nu - 0.78$ MeV, 
$R(\varepsilon_p, \varepsilon'_{p})$ is the detector response assumed to be the
Gaussian distribution of the energy resolution of $6.4 \% / \sqrt{\varepsilon'_p \, {\rm (MeV)}}$.
The neutrino event rate in the energy range $\varepsilon_{pL} \le \varepsilon_p \le \varepsilon_{pU}$
is obtained using
\begin{equation}
\frac{dN(t; \varepsilon_{pL}:\varepsilon_{pU})}{dt} = 
\int_{\varepsilon_{pL}}^{\varepsilon_{pU}} 
\frac{d^2 N(t,\varepsilon_{p})}{dt d\varepsilon_{p}} d\varepsilon_{p}, 
\end{equation}
where $\varepsilon_{pL}$ and $\varepsilon_{pU}$ are the lower and upper limits of the event energy.

\begin{table*}
\caption{The $p(\bar{\nu}_e,e^+)n$ events integrated for seven days before SN explosions
by current and future neutrino observatories.
}
\begin{center}
\begin{tabular}{lcccccc}
\tableline\tableline
 & \multicolumn{2}{c}{12 $M_\odot$} & \multicolumn{2}{c}{15 $M_\odot$} & 
\multicolumn{2}{c}{20 $M_\odot$}  \\
 \cline{2-3} \cline{4-5} \cline{6-7}
Detector & Normal & Inverted & Normal & Inverted & Normal & Inverted \\
\tableline
KamLAND (Average efficiency) & 7  & 4 & 10 & 5 & 14 & 7 \\
KamLAND (No balloon) & 10 & 5 & 15 & 8 & 20 & 11 \\
SNO+ & 10  & 5 & 14 & 7 & 19 & 10 \\
Borexino & 3  & 2 & 5 & 3 & 7 & 4 \\
JUNO & 232 & 126 & 347 & 184 & 480 & 251 \\
RENO-50 & 211 & 115 & 315 & 167 & 436 & 228 \\
LENA & 585 & 318 & 874 & 464 & 1211 & 632 \\
Super-Kamiokande & 8 & 5 & 15 & 9 & 24 & 14 \\
Hyper-Kamiokande ($E_{\nu, {\rm th}}$ = 4.79 MeV) & 134 & 80 & 250 & 146 & 406 & 233 \\
Hyper-Kamiokande ($E_{\nu, {\rm th}}$ = 5.29 MeV) & 61 & 37 & 120 & 71 & 194 & 112 \\
Hyper-Kamiokande ($E_{\nu, {\rm th}}$ = 6.29 MeV) & 12 & 8 & 27 & 16 & 42 & 25 \\
Super-Kamiokande with Gd (50\% detection efficiency) & 146 & 79 & 218 & 115 & 302 & 8 \\
Hyper-Kamiokande with Gd (50\% detection efficiency) & 2466 & 1342 & 3688 & 1957 & 5151 & 2666 \\
\tableline\tableline
\end{tabular}
\end{center}
\label{tab:preSNevents}
\end{table*}

We discuss the time evolution of the expected neutrino events by KamLAND.
Here, we consider the neutrino events integrated from the last step of the evolution 
calculation to a time before the stellar collapse:
\begin{equation}
N(t_r; \varepsilon_{pL}:\varepsilon_{pU}) = 
\int_{t_f-t_r}^{t_f} \frac{dN(t'; \varepsilon_{pL}:\varepsilon_{pU})}{dt'} dt',
\end{equation}
where $t_f$ is the time at the last step and $t_r$ is equal to $t_f - t$, i.e., the period from a time
to the last-step time.
Figure \ref{fig:eventKL} shows the integrated neutrino events by KamLAND with the energy range
of $0 \le \varepsilon_p \le \infty$.
It is convenient to use this figure when we evaluate the neutrino events during a given period
to the core collapse.
We assume that the effective livetime $\epsilon_{\rm live}$ is 0.903 and 
use the average value 0.64 for the detection efficiency $\epsilon_{\rm s}$ 
\cite{Asakura16}, which we call ^^ ^^ Average efficiency".

The $p(\bar{\nu}_e,e^+)n$ events integrated for seven days before SN explosions observed by
current and future neutrino observatories are listed in Table \ref{tab:preSNevents}.
We expect the neutrino events of 7--14 in the normal mass hierarchy and
4--7 in the inverted mass hierarchy for one week before a SN explosion if the SN explode
at the distance of $\sim 200$ pc.
The smaller event number in the inverted mass hierarchy is due to the fact that almost 
all electron antineutrinos have been converted from $\mu$ or $\tau$ antineutrinos through 
the MSW effect and that the emission rate of the $\mu$ and $\tau$ antineutrinos is smaller than 
that of electron antineutrinos (see Sec. II and Sec. III).
For more massive stars, the number of the total neutrino events is larger and the period of 
observable neutrino events is shorter.
Thus, combined with these two features, we could constrain the stellar mass of a SN 
from the observations of preSN neutrinos.

We should note that only several events will be observed by KamLAND at most in the inverted mass 
hierarchy.
In this case, it will be quite difficult to constrain the stellar mass of the preSN star because of 
statistical and systematic errors of the neutrino events.
This difficulty also should be considered in the observations by other neutrino detectors.
Uncertainties of the neutrino events by the stellar evolution models will be discussed in Sec. V.

\begin{figure}[b]
\includegraphics[width=6cm,angle=270]{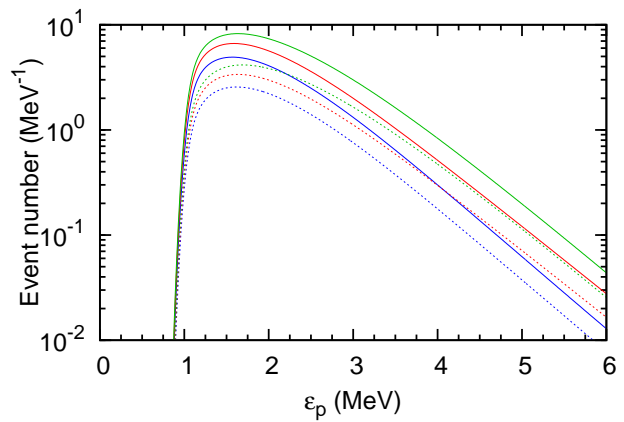}
\caption{
Expected spectrum of neutrino events detected by KamLAND.
Blue, red, and green lines indicate the 12, 15, and 20 $M_\odot$ models.
Solid and dashed lines correspond to the normal and inverted mass hierarchies, respectively.
\label{fig:spectKL}
}
\end{figure}

We show the expected spectrum of neutrino events detected by KamLAND in Fig. \ref{fig:spectKL}.
Although the average energy is slightly larger for more massive models, the stellar mass dependence 
of the spectrum feature is small.
The difference is mainly the total event number.
Owing to this reason, higher energy events may be observed for a more massive model.

At present, KamLAND contains an inner balloon for neutrinoless double beta-decay experiment
(KamLAND-Zen) \cite{Asakura16b}.
Asakura \emph{et al.} \cite{Asakura16} adopted the energy dependent efficiency 
$\varepsilon_{\rm s}(\epsilon_p)$ due to the Likelihood selection.
The main effect of the efficiency loss is the inner balloon cut \cite{Asakura16}.
The observed neutrino events depend on the detection efficiency determined by the current
installed system.
So, we also consider the efficiency without the inner balloon cut (No balloon).
In this case, the efficiency is about 0.9 \cite{Asakura16}.
This efficiency gives more neutrino events than the average efficiency.
The detection efficiency of KamLAND can increase depending on the future experiment system.

An alarm of a SN explosion using the observation of preSN neutrinos has been discussed \cite{Asakura16}.
They expected that the $3\sigma$ detection of preSN neutrinos 2--90 hours before the SN explosion
is possible with counting the neutrino events for 48 hours.
The main background for preSN neutrinos is reactor neutrinos and the background rate
is $B_{\rm low} = 0.071$ events day$^{-1}$ in the low-reactor phase and 0.355 events day$^{-1}$ 
in the high-reactor phase.
Here, we also estimate an alarm of a nearby SN explosion using the detection of three preSN neutrino
events by KamLAND for 48 hours.
Three events for 48 hours correspond to the detection significance of
$3.7 \sigma$ and $2.1 \sigma$ for low-reactor phase and high-reactor phase, respectively.
So, three events for 48 hours by KamLAND can be a good indicator of a SN alarm.
When we evaluate the time, we consider the energy range of the neutrino signal as
$0.9 \le \varepsilon_p \le 3.5$ MeV in accordance with \cite{Asakura16}.

We also consider a SN alarm in the case of No balloon.
In this case, larger efficiency also raises the background events.
We expect $B_{\rm low} = 0.103$ and $B_{\rm high} = 0.516$ using the detection efficiency 
averaged in the detected energy of 0.93.
We obtain that three events for 48 hours in low reactor phase correspond to $3.2 \sigma$ detection 
significance and that six events in high reactor phase correspond to $3.3 \sigma$.
We consider that the observations of these events for 48 hours give a SN alarm.

\begin{table}[b]
\caption{Expected SN alarm time by KamLAND.
Norm: normal mass hierarchy, Inv: inverted mass hierarchy.
}
\begin{center}
\begin{tabular}{lcccccc}
\tableline\tableline
 & \multicolumn{6}{c}{Time prior to the explosion (hours)}  \\
\cline{2-7}
 & \multicolumn{2}{c}{Average efficiency} & \multicolumn{4}{c}{No balloon} \\
 \cline{2-3}  \cline{4-7}
 &  \multicolumn{2}{c}{Three events} &  \multicolumn{2}{c}{Three events} &
  \multicolumn{2}{c}{Six events} \\
  \cline{2-3} \cline{4-5} \cline{6-7}
Model &   Norm & Inv & Norm & Inv & Norm & Inv \\
\tableline
12 $M_\odot$ & 3.5  & --- & 17.2 & 0.53 & 0.18 & --- \\
15 $M_\odot$ & 18.1  & 0.94 & 22.9 & 8.6 & 6.1 & 0.011 \\
20 $M_\odot$ & 9.4  & 3.6 & 11.9 & 7.1 & 7.0 & 0.93 \\
\tableline\tableline
\end{tabular}
\end{center}
\label{tab:KL48}
\end{table}

Table \ref{tab:KL48} shows the expected SN alarm time, i.e., the time when three neutrino events are 
expected in 48 hours.
When we consider averaged efficiency case, the expected time for the SN alarm is 
3.5--18.1 hours and less than 3.6 hours in the normal and inverted mass hierarchies, respectively.
In the normal mass hierarchy, the alarm may be provided just after the termination of the Si core burning.
The 15 $M_\odot$ model gives the longest time prior to the explosion.
This is due to moderate Si-burning period and moderate neutrino emission.
In the 12 $M_\odot$ model, the period of the Si burning is longer but the neutrino emission rate is low, 
so neutrino event number within a given period increases less steeply than the 15 $M_\odot$ model.
In the inverted mass hierarchy, it may be difficult to send a SN neutrino alarm because of less
neutrino events.
In the 20 $M_\odot$ model, the short period prior to the explosion is mainly due to the short Si-burning
period.

When we consider no balloon case, three and six events correspond to low and high reactor phases,
respectively.
In low reactor phase, the expected SN alarm time is earlier than the average efficiency case
owing to the larger detection efficiency.
The difference is larger for the SN alarm from a less massive star.
Even in high reactor phase, the SN alarm several hours before the explosion is possible for
the SN of more massive than a $\sim 15 M_\odot$ star and in the normal mass hierarchy.
The SN alarm using preSN neutrinos will extend the possibility of the SuperNova
Early Warning System (SNEWS) \cite{Antonioli04}.

\subsection{SNO+ and Borexino}

SNO+ is a neutrino experiment in SNOWLAB, Sudbury, Canada (recent review: \cite{Andringa15}).
This experiment is planned to detect neutrinos using 780 tons of liquid scintillator.
The main target is a search for the neutrinoless double-beta decays of $^{130}$Te, and
other broad topics of neutrino experiments will be performed.
The detection of SN neutrinos is one topic of the experiments.
SN neutrinos are mainly observed through $p(\bar{\nu}_e,e^+)n$ as a prompt signal
and $n(p,\gamma)d$ as a delayed signal.
The experimental facility of SNO+ such as the vessel volume and the use of liquid scintillator is
similar to KamLAND.
Table \ref{tab:preSNevents} shows the expected preSN neutrino events for seven days before the SN explosion
in SNO+.
Here, we assume 780 ton fiducial mass and the detection efficiency of 
$\varepsilon_{\rm live} = \varepsilon_{\rm S}(\varepsilon_p) = 1$ for simplicity.
These event numbers are scaled proportionally to the fiducial volume of KamLAND
with the same assumptions.
So, the preSN neutrino events will be smaller than the cases of KamLAND with no balloon.

We expect that SNO+ can also give a SN alarm using preSN neutrinos.
In this case, the background is mainly determined by reactor neutrinos.
The number of reactor $\bar{\nu}_e$ events in SNO+ is expected to be around 90 events per year
\cite{Andringa15}, corresponding to 0.25 events per day.
In this background, the detection significance of three $\bar{\nu}_e$ events for 48 hours
is $2.5\sigma$.
Thus, if the detectability of preSN neutrinos by SNO+ is similar to KamLAND, SNO+ will also observe
the preSN neutrino events expected in Fig. \ref{fig:eventKL} and will give a SN alarm.
Observing preSN neutrinos by the two neutrino experiments will increase the reliability 
of the SN alarm system.

Borexino is a liquid scintillation neutrino detector in the Laboratori Nazionali del Gran Sasso, Italy.
This experiment observes low energy solar neutrinos \cite{Bellini14}
as well as geo-neutrinos \cite{Agostini15}.
The geo-neutrino experiment observes electron antineutrinos through $p(\bar{\nu}_e,e^+)n$
similar to KamLAND, so preSN neutrinos will be observed through this reaction.
This experiment uses 278 tons liquid scintillator and the target proton number is 
$\sim 1.7 \times 10^{31}$ \cite{Bellini10}.
This proton number corresponds to about one third of KamLAND, so 
the number of preSN neutrino events is also expected to be one third of KamLAND.
Thus, 2--7 preSN neutrino events are expected to be observed in Borexino if the detection efficiency
is one (see Table \ref{tab:preSNevents}).

\begin{figure*}
\includegraphics[width=6cm,angle=270]{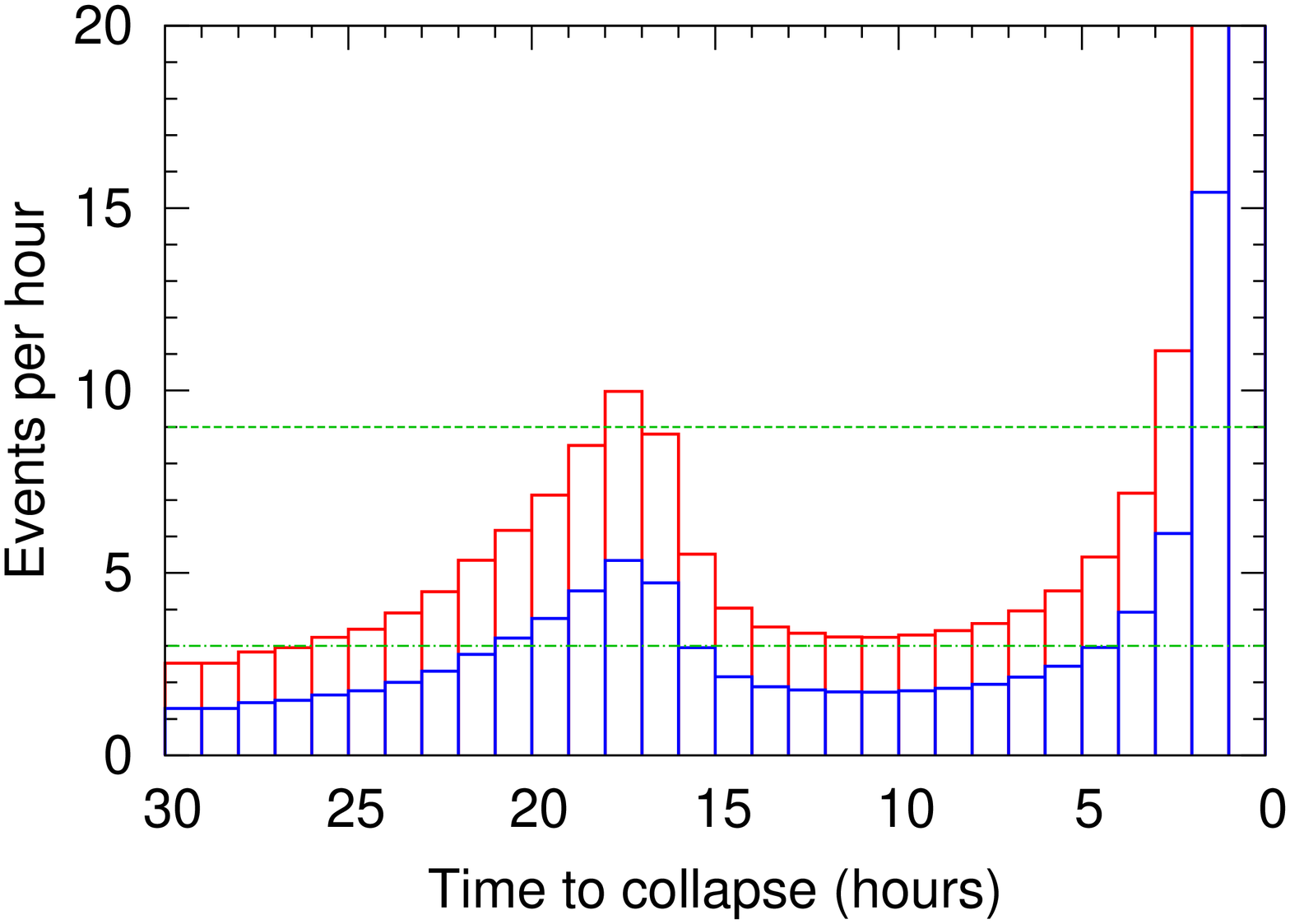}
\includegraphics[width=6cm,angle=270]{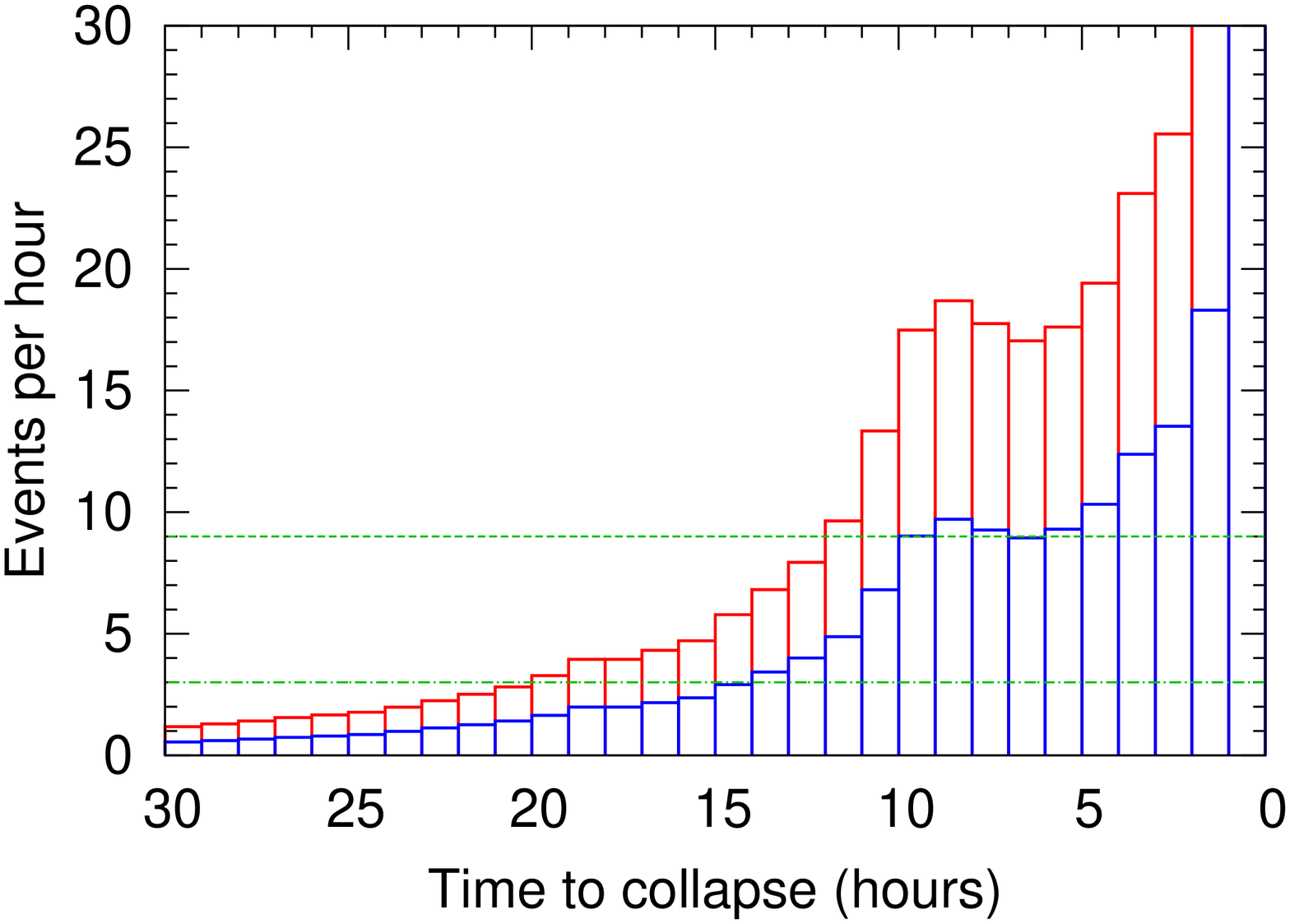}
\caption{
Expected neutrino events per one hour detected by JUNO in 30 hours prior to a SN explosion.
Left panel shows the 15 $M_\odot$ model and right panel shows the 20 $M_\odot$ model.
Red and blue lines indicate the cases of the normal and inverted mass hierarchies.
Dashed and dash-dotted green lines indicate the background events per hour 
in high and low reactor phases.
\label{fig:num60}
}
\end{figure*}

\subsection{JUNO, RENO-50, and LENA}

There are planned and proposed neutrino experiments using large size liquid scintillation detectors.
JUNO in China \cite{An15} and RENO-50 in Korea \cite{Seo15} have plans for constructing 20 
and 18 kton size detectors, respectively.
LENA in Europe has a plan of the construction of a 50 kton size detector
\cite{Wurm12}.
These detectors will enable to raise the neutrino detectability owing to the large size
comparably to the Super-Kamiokande and the low energy threshold similar to KamLAND.
If these neutrino experiments operate, the observed events of preSN neutrinos
will increase drastically.

Here, we will evaluate preSN neutrino events by these detectors based on the current proposals
of JUNO \cite{An15}.
JUNO observes electron antineutrinos through $p(\bar{\nu}_e,e^+)n$ as a prompt reaction
and $n(p,\gamma)d$ as a delayed reaction similar to KamLAND.
The fiducial mass is 20 kton and the corresponding target proton number is $1.5 \times 10^{33}$.
The detection efficiency from the fiducial mass to reduce the background is 0.79
(see Table 2-1 in \cite{An15}).
In this case, the event number is expected to be 34 times as large as in KamLAND with average
detection efficiency.

When reactor neutrino experiments are conducted in JUNO, neutrinos from reactors become background
against preSN neutrinos.
The background during reactor neutrino experiments is estimated to be 60 events per day \cite{An15}.
When the reactors are turned off, the background is estimated to be 3.8 events per day \cite{An15}.
So, we consider that the background neutrino events are 63.8 and 3.8 for high- and low-reactor phases.
These events correspond to 2.66 and 0.16 events per hour.
The number of neutrino events with the significance more than $3\sigma$ is nine and three.

The event number of preSN neutrinos by JUNO is estimated 
from the result of KamLAND (see Fig. \ref{fig:eventKL}) and the ratio of the target proton numbers.
We show the event number by JUNO for seven days before the SN explosion in Table \ref{tab:preSNevents}.
The event number by RENO-50 and LENA is also calculated using the ratio of the fiducial mass
and is listed in Table \ref{tab:preSNevents}.
We expect hundreds of preSN neutrino events will be observed by these neutrino observatories.

We show the expected neutrino events per one hour detected by JUNO in 30 hours prior to the
SN explosions of the 15 and 20 $M_\odot$ models in Fig. \ref{fig:num60}.
We see a minimum of the neutrino events around ten hours prior to the explosion
in the 15 $M_\odot$ model.
The minimum value is less than five and the maximum event number prior to the minimum is
about ten.
This minimum corresponds to the decrease in the high energy neutrino emission due to the ignition of 
the O shell burning after the termination of the Si core burning.
In the case of the 12 $M_\odot$ model, there is a peak with eight (four) events in the normal (inverted) 
mass hierarchy at about sixteen hours before the explosion.
The event number per hour becomes a minimum around ten hours before the explosion.
Since the peak height is smaller than the 15 $M_\odot$ model, it is more difficult to observe it.

\begin{figure}
\includegraphics[width=6cm,angle=270]{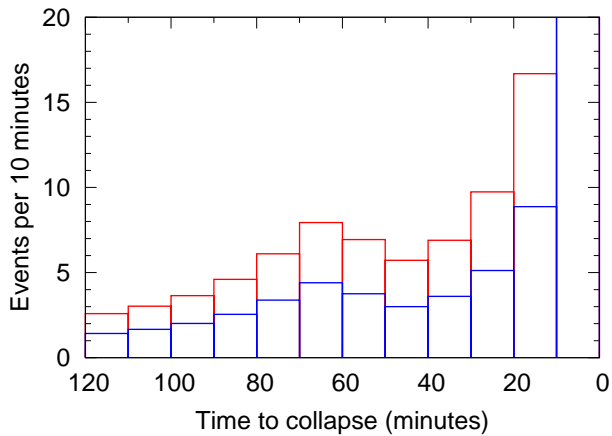}
\caption{
Expected neutrino events per ten minutes detected by JUNO in 120 minutes prior to a SN explosion
of the 15 $M_\odot$ model.
Red and blue lines indicate the cases of the normal and inverted mass hierarchies.
\label{fig:num10}
}
\end{figure}

We also see a minimum of the neutrino events around seven hours prior to the explosion 
in the 20 $M_\odot$ model.
This change is less prominent than the 15 $M_\odot$ model but still we would recognize
the change of burning processes in the central region.
The 20 $M_\odot$ model indicates weak neutrino emission from the O shell.
Thus, we could observe the evolution of burning processes in the central region of collapsing stars
through the preSN neutrino events and could constrain shell burnings after the Si core burning.

We present the time evolution of the neutrino events just before the SN explosion.
Figure \ref{fig:num10} shows the expected neutrino events for ten minutes detected by JUNO
for two hours prior to the explosion in the 15 $M_\odot$ model.
We see a minimum of the event around 50 minutes before the explosion but the difference of
the event number at the maximum around 70 minutes is small.
This corresponds to the ignition of the Si shell burning at $M_r \sim 1 M_\odot$.
This signal also would be observed by Hyper-Kamiokande if low threshold energy is set.
The multiple observations of preSN neutrinos will raise the reliability of observed neutrino signals.
We note that the neutrino events per ten minutes monotonically increase for two hours
in the 20 $M_\odot$ model.
Since the convection of the Si shell burning of the 20 $M_\odot$ model is not so strong,
the efficiency of the convection would be constrained from the preSN neutrino signals.

\begin{figure*}
\includegraphics[width=6cm,angle=270]{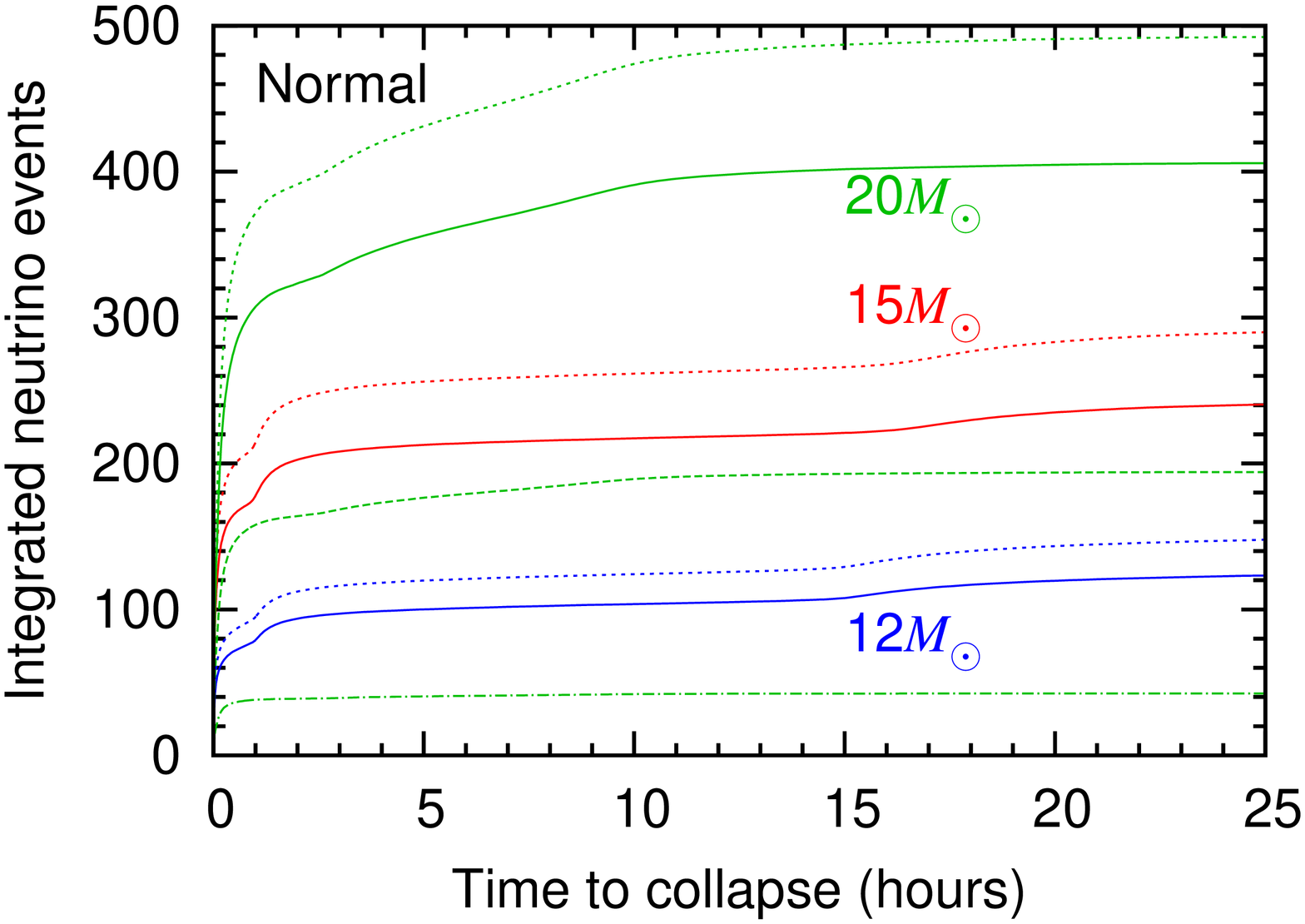}
\includegraphics[width=6cm,angle=270]{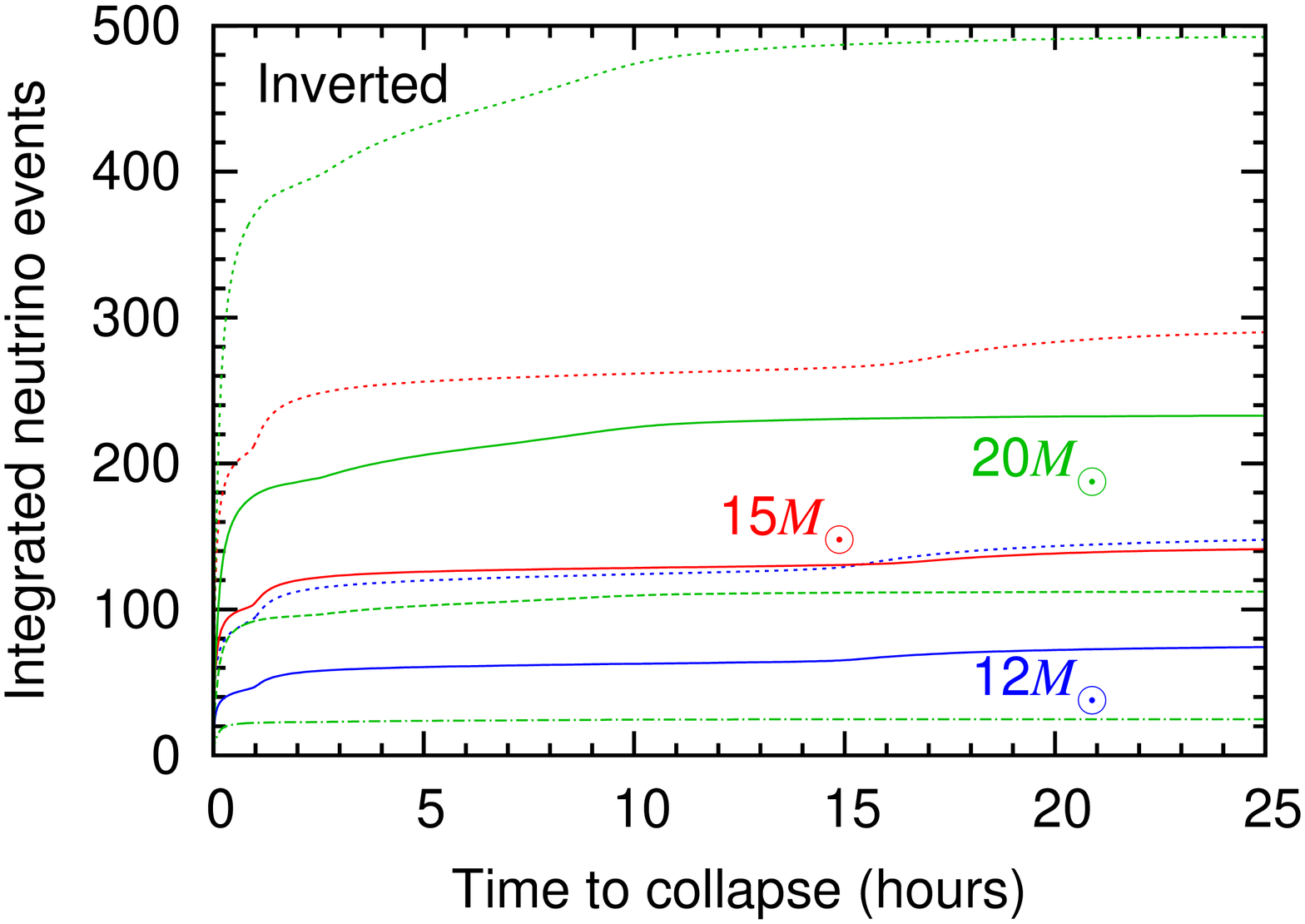}
\caption{
Same as Fig. \ref{fig:eventKL} but for the detection by Hyper-Kamiokande with the neutrino threshold 
energy of 4.79 MeV.
Dashed and dash-dotted green lines indicate the neutrino events with the neutrino threshold energy of 
5.29 and 6.29 MeV, respectively.
\label{fig:eventHK}
}
\end{figure*}

A SN alarm using preSN neutrinos is also possible for JUNO.
We consider the time for the SN alarm using the neutrino events per hour of the three models.
The energy range for the alarm is set to be $0.9 \le \varepsilon_p \le 3.5$ MeV similar to KamLAND.
The energy resolution is assumed to be the same as KamLAND.
Table \ref{tab:JN1} lists the expected SN alarm time given by JUNO.
In the low reactor phase, the SN alarm will be sent earlier than KamLAND.
Except for the case of the inverted mass hierarchy of the 12 $M_\odot$ model,
the alarm will be sent before the star starts the O shell burning.

\begin{table}[b]
\caption{Expected SN alarm time by JUNO (hours prior to the explosion).
}
\begin{center}
\begin{tabular}{ccccc}
\tableline\tableline
 & \multicolumn{4}{c}{Time prior to the explosion (hr)}  \\
 \cline{2-5}
 & \multicolumn{2}{c}{Low reactor phase} & \multicolumn{2}{c}{High reactor phase}  \\
 \cline{2-3} \cline{4-5}
Model &   normal & inverted & normal & inverted\\
\tableline
12 $M_\odot$ & 19.9  & 16.1  & 1.29 & 0.640 \\
15 $M_\odot$ & 23.6  & 20.0 & 17.2 & 1.31 \\
20 $M_\odot$ & 16.8  & 12.6 & 10.9 & 8.46 \\
\tableline\tableline
\end{tabular}
\end{center}
\label{tab:JN1}
\end{table}

We note that one of the main targets in JUNO and RENO-50 experiments is reactor neutrino experiment.
The neutrino background will be high during the reactor neutrino experiment.
Since preSN neutrinos are hidden in the high background, the identification and sending a SN alarm
will be delayed.
In high-reactor phase, it may be difficult to observe a minimum of the neutrino events
after the Si core burning by JUNO (see Fig. \ref{fig:num60}).
The SN alarm by JUNO is delayed to KamLAND in the low reactor phase in some cases.
Although the power of nuclear plants for RENO-50 is smaller than JUNO, the SN alarm time is still similar
to KamLAND or worse.
So, we consider that monitoring preSN neutrinos and sending an alarm by KamLAND and SNO+ 
are also important.
In order to send a SN alarm in an early time before the SN explosion, monitoring by many
neutrino detectors with a low energy threshold is desirable.

\subsection{Super-Kamiokande and Hyper-Kamiokande}

Super-Kamiokande is a water Cherenkov detector located in the Kamioka Mine, Japan.
In the fourth phase of solar neutrino experiment, the fiducial volume is 22.5 kton for most of 
the energy range and the threshold energy of the recoil electrons reduces to 3.5 MeV \cite{Sekiya13}.
Supernova neutrinos will be detected mainly by $p(\bar{\nu}_e,e^+)n$ reaction.
In this case, the observable threshold for neutrinos is 4.79 MeV.
Hyper-Kamiokande is proposed as a next generation water Cherenkov detector.
The proposed fiducial volume has been recently changed to 380 kton.
The threshold energy is expected to be lower than the previous plan \cite{AbeK11a}.
Here, we estimate the neutrino events of Super-Kamiokande and Hyper-Kamiokande.
We assume for Hyper-Kamiokande that the fiducial volume is 380 kton and 
the neutrino threshold energy is 4.79 MeV.
The neutrino events of Super-Kamiokande can be estimated by scaling with a factor 22.5/380 = 0.06.

\begin{figure}[b]
\includegraphics[width=6cm,angle=270]{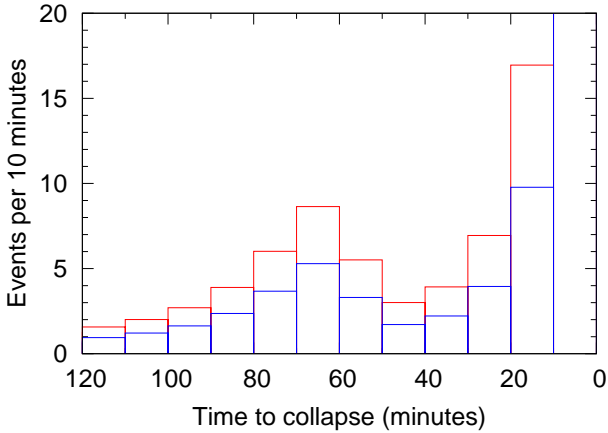}
\caption{
Same as Fig. \ref{fig:num10} for the detection by Hyper-Kamiokande 
with the neutrino energy threshold of 4.79 MeV.
\label{fig:num10_HKn4}
}
\end{figure}

We estimate the integrated $\bar{\nu}_e$ events by Hyper-Kamiokande in Fig. \ref{fig:eventHK}.
The neutrino events for seven days before the explosion by Super-Kamiokande and Hyper-Kamiokande
are listed in Table \ref{tab:preSNevents}.
Most of the preSN neutrino events will be observed in one day before the explosion.
Although the threshold energy is higher than liquid-scintillation observatories, 
the large fiducial volume makes it possible to observe hundreds of neutrino events.
In Super-Kamiokande, several to tens of neutrino events will be observed.

\begin{figure*}
\includegraphics[width=6cm,angle=270]{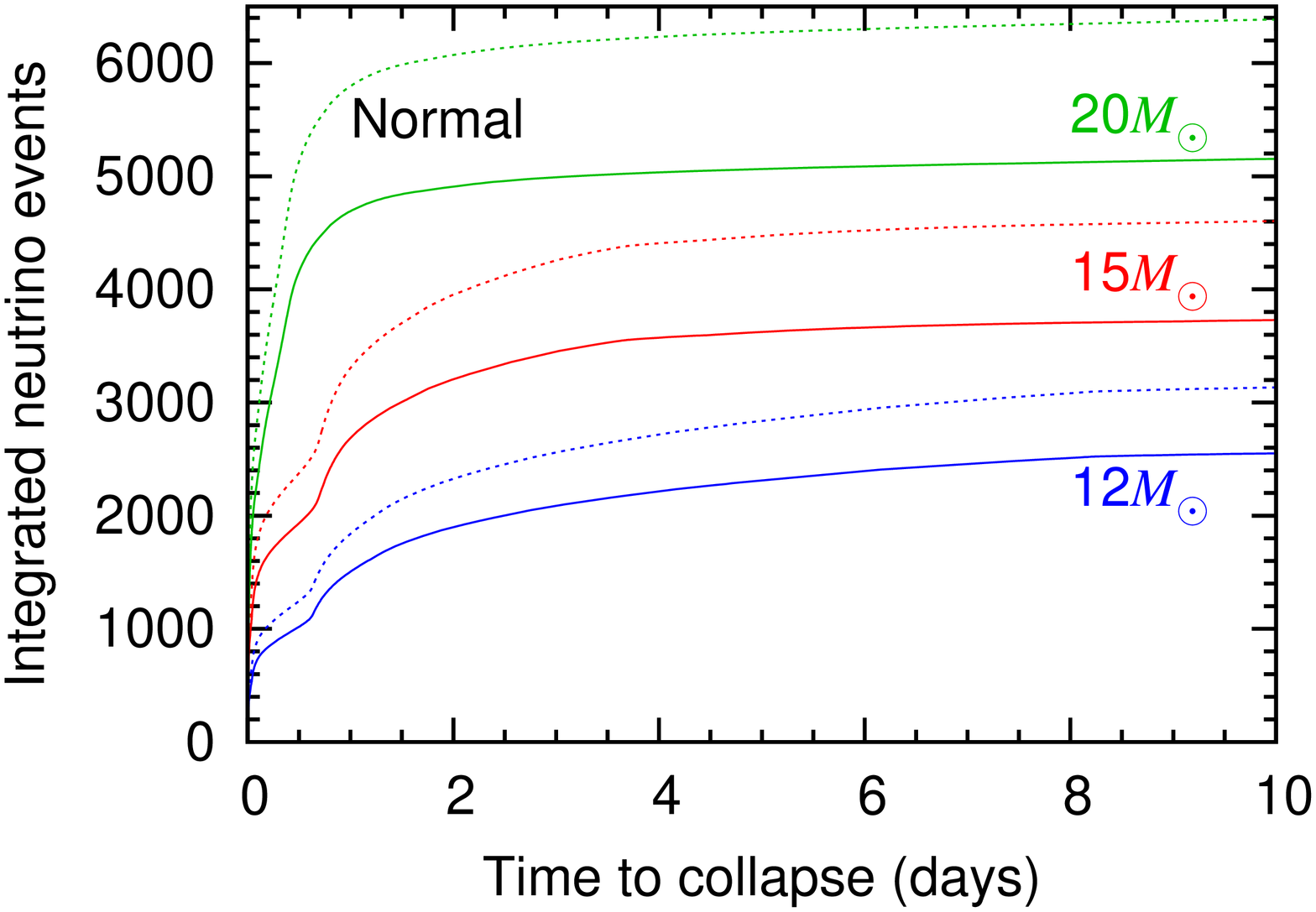}
\includegraphics[width=6cm,angle=270]{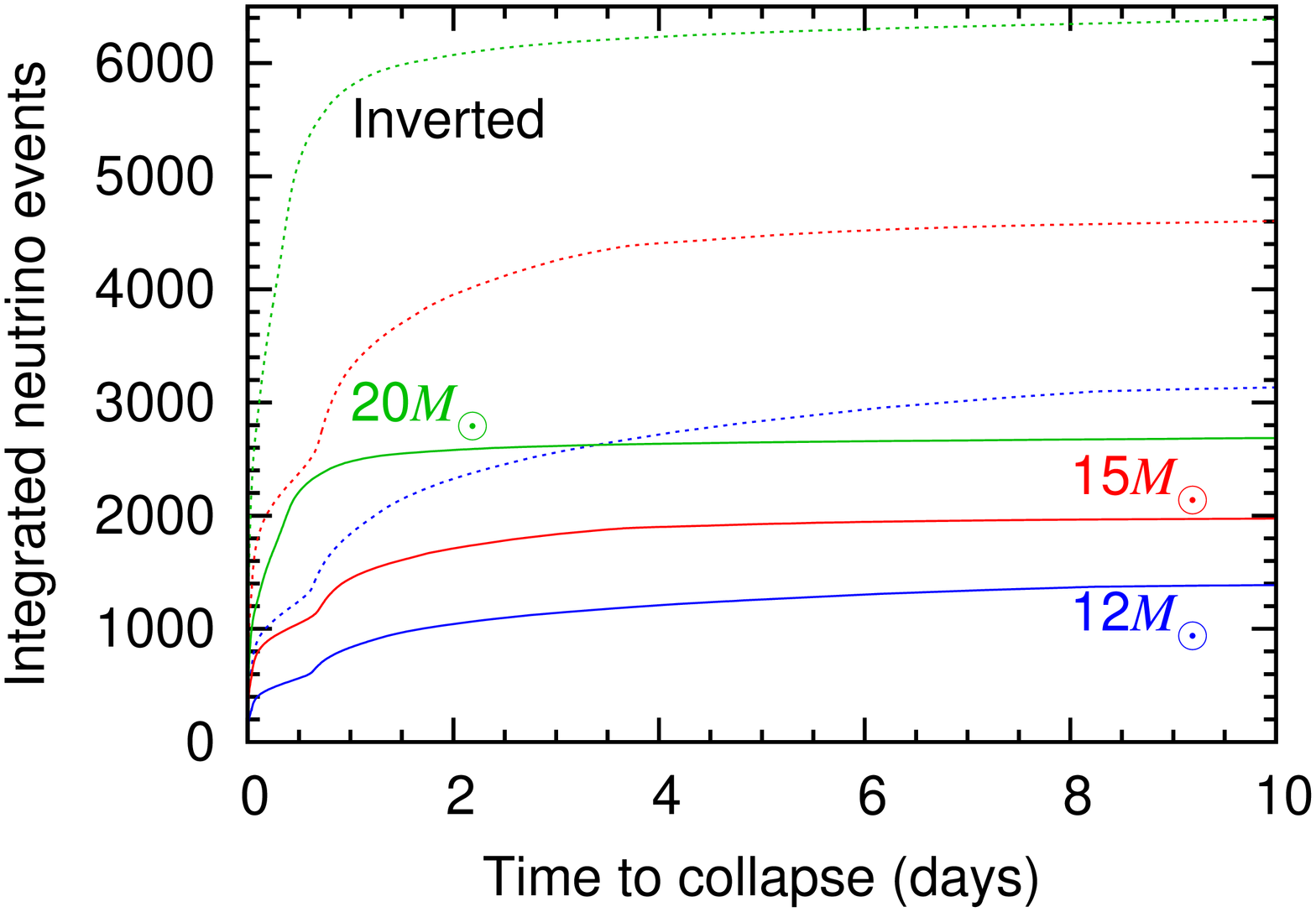}
\caption{
Same as Fig. \ref{fig:eventKL} but for the detection using delayed signals by Gd-loaded Hyper-Kamiokande. \label{fig:eventHK0}
}
\end{figure*}

\begin{figure}
\includegraphics[width=6cm,angle=270]{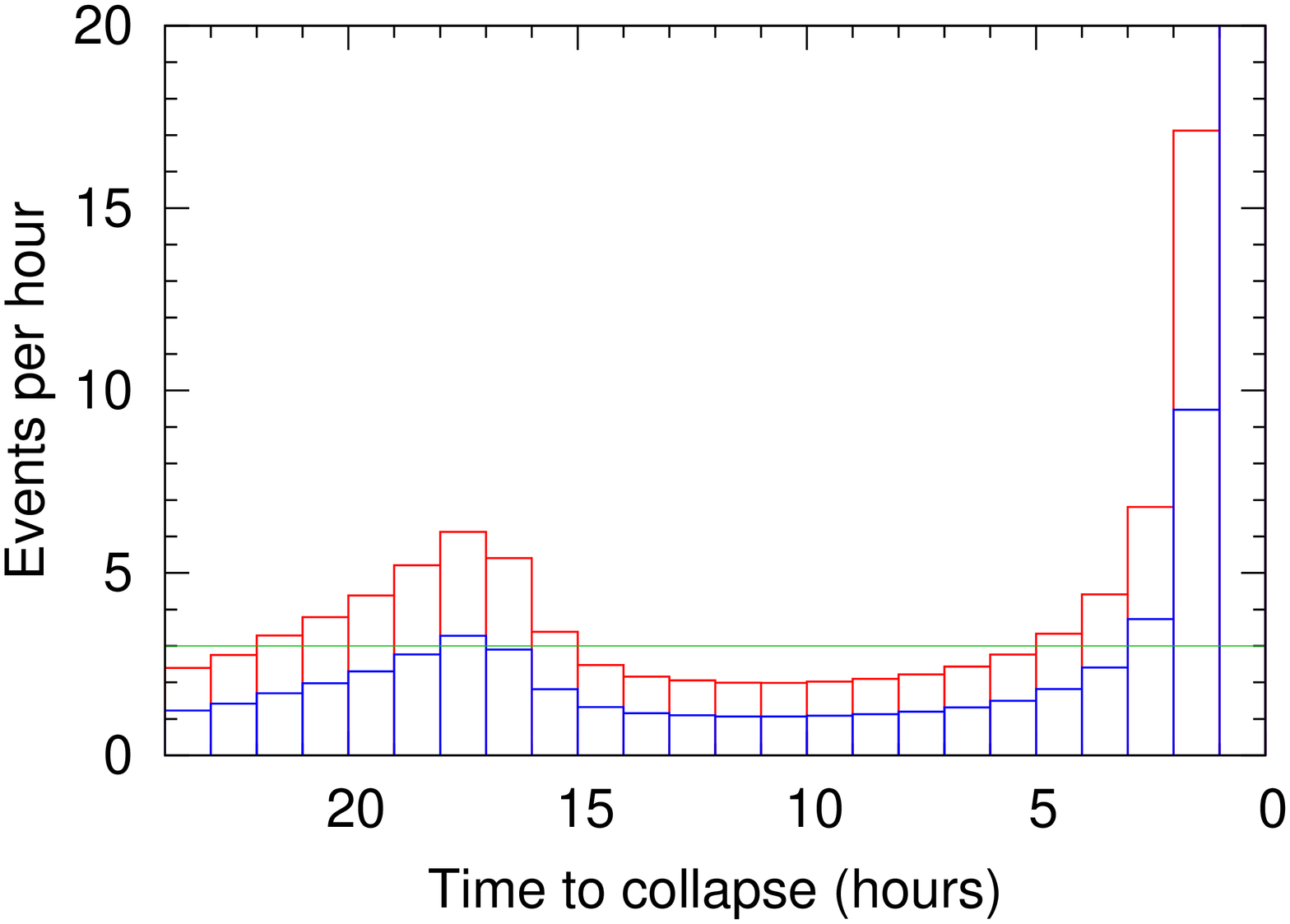}
\includegraphics[width=6cm,angle=270]{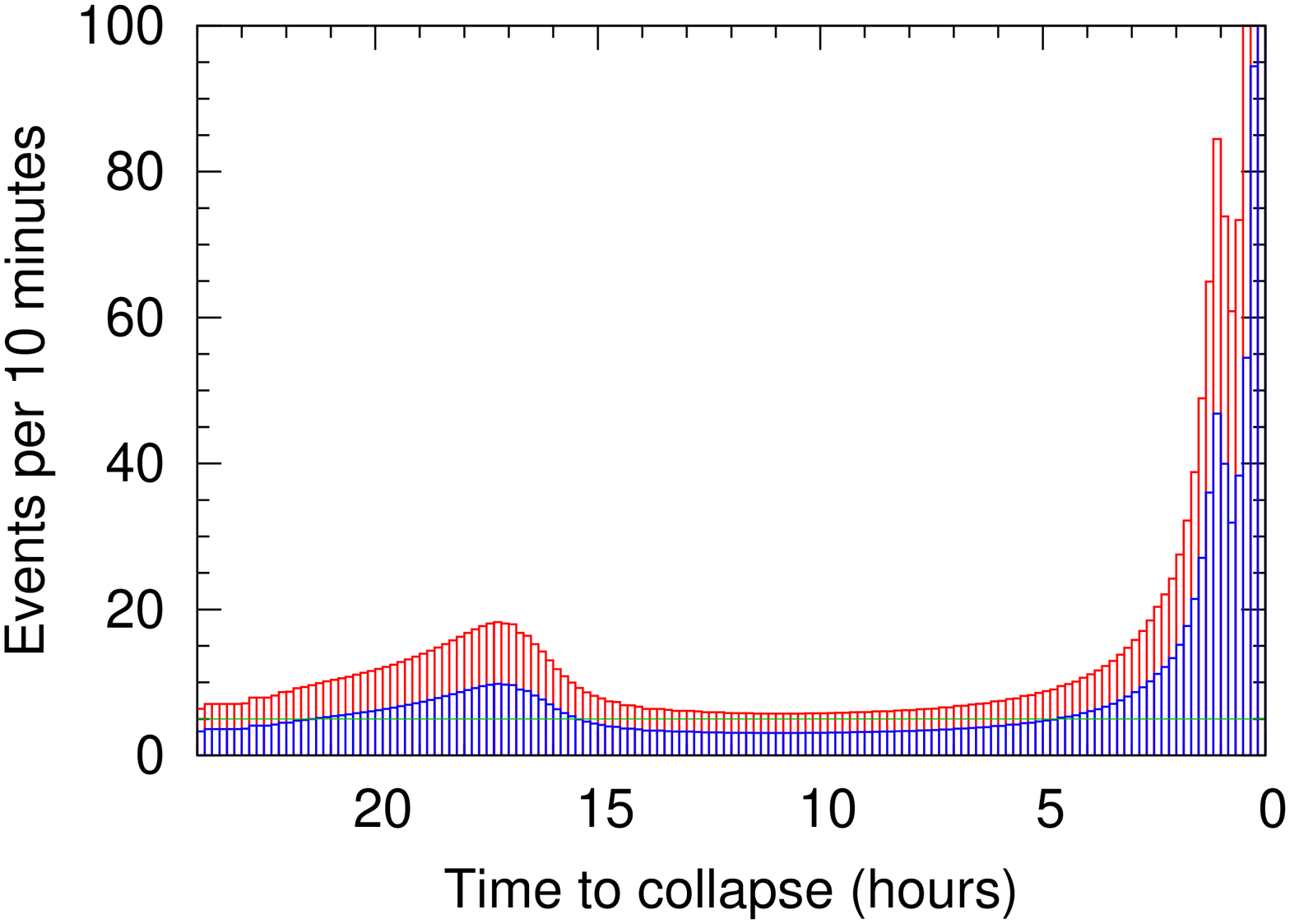}
\caption{
Expected neutrino events per hour detected using delayed signals by Super-Kamiokande with Gd
(top panel) and neutrino events per 10 minutes by Hyper-Kamiokande with Gd
(bottom panel) in 24 hours prior to a SN explosion of the 15 $M_\odot$ model.
Red and blue bins indicate the cases of the normal and inverted mass hierarchies.
The green horizontal line is the event number for the unit interval of the significance 
more than $3\sigma$ (see text for details).
\label{fig:num10_m15HKn0b}
}
\end{figure}

The expected neutrino events strongly depend on the threshold energy.
We also show the integrated $\bar{\nu}_e$ events for the 20 $M_\odot$ model 
in $E_{\nu,{\rm th}}$ = 5.29 and 6.29 MeV in Fig. \ref{fig:eventHK}.
The neutrino event number is smaller by factors of about two and ten.
Thus, it is very important for the observation of preSN neutrino events to achieve low energy threshold.

Owing to the large fiducial volume, Hyper-Kamiokande also would observe the change of burning
processes during the final stage of massive star evolution.
Figure \ref{fig:num10_HKn4} shows the expected $\bar{\nu}_e$ events per ten minutes of 
the 15 $M_\odot$ model by Hyper-Kamiokande.
We see the decrease in the neutrino events before about one hour to the explosion and 
the observed neutrino events are similar to JUNO (see Fig. \ref{fig:num10}).
Observations of different types of neutrino detectors will increase the reliability of the neutrino events
caused by processes during the massive star evolution.

\subsection{Gd-loaded Super-Kamiokande and Hyper-Kamiokande}

The Gadzooks! project is an establishment of the identification of $p(\bar{\nu}_e,e^+)n$ event by 
neutron tagging by Gd \cite{Beacom04}.
When small amount of Gd is contained in water Cherenkov detector, neutrons produced through
the inverse beta-decay are captured by Gd and $\gamma$-rays with $\sim 8$ MeV are released.
The prompt signal by $e^-e^+$-annihilation tagged by the $\gamma$-rays by the neutron capture
is identified as this event.
In preSN neutrinos, since the average $\bar{\nu}_e$ energy is below the observation threshold,
most of the prompt signal will not be detected.
However, the delayed signals are detectable.
If many {\it delayed} signals are detected, these signals will be recognized as preSN neutrinos
{\it even after} the corresponding SN explosion.
We estimate preSN neutrino events detected by the delayed signal in Gd-loaded 
Super-Kamiokande and Hyper-Kamiokande.

We assume the detection efficiency of 0.5 for Gd-loaded Super-Kamiokande and Hyper-Kamiokande.
This efficiency roughly corresponds to the detection of lower energy region in the energy spectrum of
the delayed $\gamma$-ray signals, since the $\gamma$-rays have the energy spectrum peaked at 
$\sim 5$ MeV \cite{Watanabe09}.
The background events are considered to be the events below 5 MeV in the third phase
solar-neutrino experiment \cite{AbeK11b}.
The average background events are 0.21 and 3.61 per hour for Super-Kamiokande and
Hyper-Kamiokande, respectively.
The event number of more than $3\sigma$ significance is three and eleven per hour.
We note that the assumption on the detection efficiency and the background is rough.
The threshold for the neutrino energy is determined by $p(\bar{\nu}_e,e^+)n$ to be 1.8 MeV.

Figure \ref{fig:eventHK0} shows the integrated neutrino events by Hyper-Kamiokande with Gd.
The neutrino events for seven days before the explosion in Super-Kamiokande and Hyper-Kamiokande 
are also listed in Table \ref{tab:preSNevents}.
Large fiducial volume and low energy threshold give many events.
We assume here that the distance to the SN is 200 pc.
Since the neutrino event number declines at inverse square of the distance, more than ten events
are expected in the distance of $\sim 3$ kpc for Hyper-Kamiokande.
Thus, the detectable region of preSN neutrino events increases by Hyper-Kamiokande with Gd.

We also investigate detailed time evolution of the preSN neutrino events.
Figure \ref{fig:num10_m15HKn0b} shows the neutrino events of the 15 $M_\odot$ model.
We investigate neutrino events per one hour for Super-Kamiokande and per ten minutes for
Hyper-Kamiokande.
We see a peak around 17 hours before the explosion for Super-Kamiokande.
The peak is expected to be detectable in the normal mass hierarchy.
The neutrino events could be identified even in the inverted mass hierarchy.
Combined with the observations by other neutrino detectors such as JUNO, we will confirm 
the time evolution of the neutrino events.
In Hyper-Kamiokande, we could see a peak at 17 hours before the explosion more clearly and 
another peak around one hour before the explosion.
These peaks appear just before the ignitions of the O shell burning and the Si shell burning.
They are above $3\sigma$ background events even in the inverted mass hierarchy.
Thus, the time evolution of the neutrino events could constrain burning processes during the final
evolution of massive stars.

The observation using delayed neutrino signals by Hyper-Kamiokande with Gd is a powerful tool
to observe preSN neutrinos.
With this method, thousands of preSN neutrino events are expected.
Although we do not show the SN alarm by Hyper-Kamiokande with Gd because of rough estimation
of the background, we expect that the SN alarm earlier than JUNO and KamLAND is possible.
After the investigation of the signal to noise ratio in Hyper-Kamiokande, we will discuss the possibility
of long period preSN neutrino observations.

\subsection{DUNE}

Deep Underground Neutrino Experiment (DUNE) is a neutrino experiment 
proposed in the United States \cite{Acciari16}.
The main characteristic of this experiment is observing electron neutrinos using a massive liquid argon
time-projection chamber.
Four detectors with the fiducial mass of 10 kton will be constructed until around 2028, so 
the total fiducial mass is planned to be 40 kton.
DUNE detector observes charged-current (CC) and neutral-current reactions of $^{40}$Ar
and electron scatterings.
A charged current reaction $^{40}$Ar($\nu_e,e^-)^{40}$K$^*$ has the threshold of 1.5 MeV and
its cross section is the largest among the reactions.

We evaluate the expected electron neutrino events through the CC reaction of $^{40}$Ar by DUNE.
We assume that the fiducial mass of the detector is 40 kton and the threshold of the neutrino 
energy is 5 MeV.
The cross section of $^{40}$Ar($\nu_e,e^-)^{40}$K$^*$ is adopted from 
\cite{Kolbe03} as numerical data in 
SNOwGLoBES \footnote{http://www.phy.duke.edu/~schol/snowglobes/} 
(see also a software package GLoBES \cite{Huber07}).
The transition probability of $\nu_e \rightarrow \nu_e$ is evaluated as
\begin{equation}
P_{(\nu_{e} \rightarrow \nu_{e})} = \left\{ 
 \begin{array}{ll}
\sin^2\theta_{13} = 0.024 & ({\rm normal}) \\
\sin^2\theta_{12} \cos^2\theta_{13} = 0.301 & ({\rm inverted}). 
 \end{array} 
 \right.
\end{equation}
The expected neutrino events are 0.11--0.32 (0.16--0.48) for one day prior to the explosion
in the normal (inverted) mass hierarchy.
The event number from seven days to the previous day of the explosion is much smaller.
Thus, we consider that it is quite difficult to observe the pair neutrino events by DUNE.
However, we should note that electron neutrino events through electron captures of nuclei
are not taken into account in this study.
The neutrino luminosity by weak interactions by nuclei becomes
larger than that by the pair neutrinos five minutes prior to the explosion in the 15 $M_\odot$ model
(see Fig. \ref{fig:lnu}).
If the average neutrino energy is higher, the observed neutrino events will be much larger.
We will investigate the neutrino events by weak interactions of nuclei in future study.

\section{Discussion}

\subsection{Neutrino events from Betelgeuse}

Betelgeuse is one of our neighboring red supergiants.
The bolometric luminosity of this star is  $\log L/L_\odot = 5.10 \pm 0.22$ \cite{Harper08} and
the effective temperature is $3641 \pm 53$ K \cite{Perrin04}.
The evolution of nonrotating and rotating massive stars like Betelgeuse has recently been 
discussed \cite{Meynet13}.

The distance to Betelgeuse has been deduced as $197 \pm 45$ pc combined with their
very large array (VLA) radio positions, the published VLA positions, and the Hipparcos
Intermediate Astrometric data \cite{Harper08}.
The distance has a large uncertainty and the uncertainty relates to uncertainties on 
the stellar luminosity and, thus, the stellar mass.
In the cases of the shortest (152 pc) and longest (242 pc) distance, the stellar luminosity 
$\log L/L_\odot$ is estimated as 4.87 and 5.28, respectively.
Compared with the final luminosity of stellar evolution models, we can estimate the initial mass range 
of the stellar evolution model (see Table \ref{tab:stars}).
In our models, the luminosity of 5.10 corresponds to about a 17 $M_\odot$ star.
The minimum and maximum luminosities correspond to about 13 and 20 $M_\odot$ stars,
respectively.

If the distance to Betelgeuse is 200 pc, the initial stellar mass is expected to be 17 $M_\odot$
in our models.
The total neutrino events will be larger and the duration of the neutrino emission will be shorter
than the case of the 15 $M_\odot$ model.
A SN alarm using preSN neutrinos could be sent within 24 hours prior to the explosion 
by KamLAND, SNO+, JUNO, and RENO-50.
The evolution of shell burnings after the Si core burning could be observed by JUNO and RENO-50
if the mass hierarchy is normal.

If the distance to Betelgeuse is $\sim 150$ pc, the neutrino flux is 1.8 times as large as the case of
200 pc.
On the other hand, the initial stellar mass is expected to be $\sim 13 M_\odot$.
From the integrated neutrino events of the three models,
we consider that the total neutrino events could be larger than 
the case of a $\sim 20 M_\odot$ star at 200 pc.
The estimation of the SN alarm time is difficult because the period of the Si core burning is longer
for less massive stars.
When the neutrino flux is 1.8 times larger, the alarm time for the 12 and 15 $M_\odot$ models
by KamLAND in the low-reactor phase is 21.7 (2.6) and 27.6 (17.1) hours before the explosion
in the normal (inverted) mass hierarchy.
We expect from the estimated alarm time that the SN alarm may be sent 
from KamLAND and SNO+ before the O shell burning starts even in the 13 $M_\odot$ star 
in the case of the normal mass hierarchy.
The closer distance also raises the possibility of the observations of the time evolution after
the Si core burning.
The evolution of shell burnings after the Si core burning could be observed more clearly
by JUNO and RENO-50.
Combined analysis of JUNO, RENO-50, and Hyper-Kamiokande could reveal evidence for the
Si shell burning within one hour before the explosion.
If Hyper-Kamiokande with Gd detects more preSN neutrino events, these events give constraints
of burning processes from the Si core burning.

If the distance to Betelgeuse is $\sim 250$ pc, the neutrino flux is 0.64 times as large as the case of
200 pc, although the initial stellar mass is expected to be $\sim 20 M_\odot$.
In this case, the expected neutrino events by KamLAND are 7 and 3 in the normal and inverted
mass hierarchies, respectively.
So, the preSN neutrino events will be observed by KamLAND and SNO+ 
even in the distance to Betelgeuse of $\sim 250$ pc.
Larger events could be observed by JUNO, RENO-50, and Hyper-Kamiokande with a low energy
threshold.
We see from the right panel of Fig. \ref{fig:num60} that about twelve events per hour could be observed 
at $\sim 8$ hours and, then, the event rate per hour could continue for three hours 
in the normal mass hierarchy.
So, even in this distance, we could obtain the information in the central region of the preSN star
from the neutrino events.
We should note that the delayed alarm time by KamLAND may make it difficult to observe 
burning processes by JUNO and RENO-50.
The SN alarm by KamLAND in the low-reactor phase will be sent at 5.9 (0.4) hours 
before the explosion in the normal (inverted) mass hierarchy.
In this case, the O shell burning has started and it is difficult to observe it.
If JUNO is in the low-reactor phase, it will send a SN alarm at 12.5 (10.0) hours
before the explosion in the normal (inverted) mass hierarchy.
This time is prior to a minimum of the neutrino events per hour in Fig. \ref{fig:num60}.
Thus, the evidence for the burning processes after the Si core burning may be still possible to
be observed by JUNO and RENO-50.

Although stellar evolution models explain many stellar phenomena, there are still uncertainties in 
their parameters.
The uncertainties will affect the predictions on the neutrino emission in the final stage of the evolution.
The 15 $M_\odot$ model in \cite{Kato15} used larger overshoot effect and different convection
treatment.
The CO-core mass in their model is larger than our model and, thus, the period from the Si core burning
to the core collapse is shorter.
They do not see clear decrease in the neutrino events by the O and Si shell burnings.
The 15 $M_\odot$ model in \cite{Asakura16}, originally in \cite{Woosley02}, seems to indicate
the Si burning period similar to our model.
The evidence for the Si shell burning is shown in around 1--2 hours before the explosion in the
neutrino luminosity (see Fig. 2 in \cite{Asakura16}).
A small dip of the neutrino luminosity around four hours before the explosion could be the evidence
for the O shell burning.

Even in a fixed stellar mass, the CO core mass has an uncertainty due to the material mixing
by convection and stellar rotation.
This uncertainty affects the time scale of the advanced evolution.
The efficiency of the convection also affects the radial distribution of the neutrino emission through
burning processes.
The observation of preSN neutrinos will be the first direct observation of the central region of a collapsing 
massive star.
If detailed time evolution of the neutrino events from an evolved massive star is observed, 
evolution signatures such as the period of the Si core burning and Si and/or O shell burnings
could be revealed.

We considered the MSW effect of neutrino oscillations in SN progenitors.
However, we do not consider the Earth effect of neutrino oscillations.
The time scale of preSN neutrinos is the order of day, which is much longer than SN neutrinos.
The neutrino path in the Earth changes in one day period.
We would like to take into account this effect in future study.
Neutrino forward scattering in SN neutrinos changes the neutrino flavors (collective oscillations) 
and the final neutrino spectra (e.g., \cite{Duan06}).
In preSN stage, since the neutrino flux is still much smaller than in the SN neutrinos, the effect of 
neutrino self-interaction is expected to be negligible for neutrino oscillations.
It is an interesting problem when neutrino scattering becomes important for neutrino oscillations
during the SN explosion.

Although there is a large uncertainty in the distance to Betelgeuse, we expect in this study that 
neutrino events will be observed for several to tens hours before Betelgeuse explodes as a SN.
The time of the SN alarm strongly depends on the distance, neutrino mass hierarchy, and 
the background of each neutrino detector.
If the mass hierarchy is normal and the background of KamLAND and JUNO is low, 
the burning processes just before the SN explosion could be revealed through the
preSN neutrino observations.
Hyper-Kamiokande with Gd could observe thousands preSN neutrino events using delayed
signals from the Gd+$n$ reaction despite the uncertainty of the distance.
The neutrino events also become a constraint of the burning processes.
The possibility of nearby SN explosions is expected to be one event per tens-thousand years; 
there are about ten evolved stars such as red supergiants and WR stars around a few hundred pc 
(e.g., \cite{Asakura16})
and the lifetime of the He burning is 10$^5$--10$^6$ years.
This possibility is certainly small, but we hope that we observe the interior of the evolved massive
star directly when a SN explodes in neighbors.

\subsection{Relation between the neutrino emission from preSN stars and stellar structure}

We investigated the dependence of the neutrino emission of preSN stars on the initial stellar mass
among the three models.
The neutrino events and the period from the Si core burning to the collapse phase correlate with 
the initial mass and final mass.
However, the final mass does not correlate with the initial mass for more massive stars because of
larger mass loss effect.
Further, even in the three models the Fe core mass does not correlate with the initial stellar mass.
On the other hand, structural characteristics of preSN stars for the likelihood of explosion 
have been investigated.
The compactness parameter \cite{O'Connor11} and the mass derivative at the location of
the dimensionless entropy per nucleon of $s=4$ \cite{Ertl16} are parameters
indicating a likelihood of SN explosion.
Here, we discuss whether these parameters become indicators of the neutrino emission from preSN
stars.

\begin{figure}[b]
\includegraphics[width=6cm,angle=270]{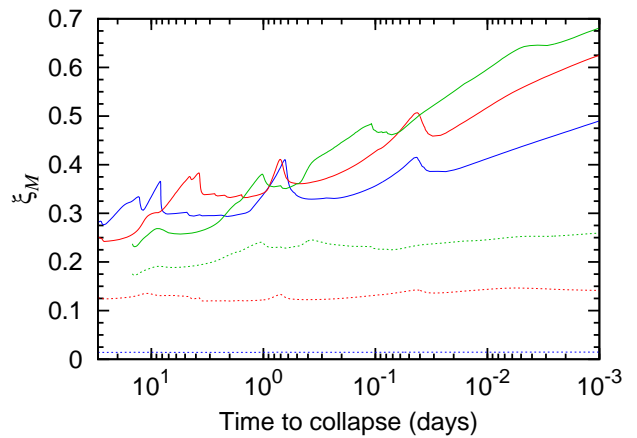}
\caption{
The time variation of the compactness parameters $\xi_{1.5}$ (solid lines) and 
$\xi_{2.5}$ (dashed lines).
Blue, red, and green lines correspond to the 12, 15, and 20 $M_\odot$ models, respectively.
\label{fig:cmpct}
}
\end{figure}

The compactness parameter $\xi_M$ is defined by
\begin{equation}
\xi_M \equiv \frac{M/M_\odot}{R(M)/1000 {\rm km}}
\end{equation}
at the core bounce, where $M$ is the specified mass coordinate and $R(M)$ is the corresponding radius.
However, we are now interested in the neutrino emission during the Si core burning and later phase.
So, we investigate the time variation of this parameter.
We consider two cases of the compactness parameter: $M$ = 1.5 and 2.5 $M_\odot$.
Figure \ref{fig:cmpct} shows the time variation of the compact parameters $\xi_{1.5}$ and $\xi_{2.5}$.
The value of $\xi_{1.5}$ increases during core contraction and decreases during core and shell burnings
among the three models.
Although the final values of the parameter have a correlation to the neutrino emission of preSN stars,
they do not reflect the structure outside the final Fe core.
On the other hand, the time variation of $\xi_{2.5}$ is much smaller and the values correlate with
the neutrino emission: larger $\xi_{2.5}$ indicates high neutrino emissivity and short time scale of
neutrino emission.
It was pointed out that $\xi_{2.5}$ might have a correlation with the period 
from the core Si depletion to the core-collapse \cite{Sukhbold14}.
So, $\xi_{2.5}$ could be an indicator characterizing the neutrino emission from preSN stars.

The relation between the normalized mass inside a dimensionless entropy per nucleon of
$s=4$, $M_4 \equiv M(s=4)/M_\odot$, and the mass derivative at this location,
$\mu_{4} \equiv (dM/M_\odot)/(dr/ 1000 \, {\rm km}) \vert _{s=4}$, was considered in \cite{Ertl16}.
The parameter $\mu_4$ is linked to the accretion rate of matter \cite{Suwa16}.
We investigated the time evolution of $M_4$ and $\mu_4$ in the three models.
In the 15 $M_\odot$ model, $M_4$ changes from the location of the outer boundary of the O/Ne layer 
to the inner boundary of the O/Ne layer by the Ne shell burning during the Si shell burning.
The change of $M_4$ during the evolution is also seen in the 20 $M_\odot$ model.
The $M_4$ value of the models before the Si core burning increases with the order of 
the 12, 20, and 15 $M_\odot$ model but it increases with the 20, 15, and 12 $M_\odot$ model 
at the last step.
The mass derivative $\mu_4$ shows a correlation with the stellar mass but the corresponding values
change with the $M_4$ value.
So, $M_4$ and $\mu_4$ indicate characteristics at the collapsing phase rather than the Si burning.
Thus, we consider that the compact parameter $\xi_{2.5}$ indicates structure characteristics from
the Si core burning to the collapsing phase and that it could characterize the neutrino emission
from preSN stars.
PreSN stars with larger $\xi_{2.5}$ will emit more neutrinos with shorter time scale of the Si burning.

\section{Conclusions}

We investigated the neutrino emission of preSN stars with the initial mass of 12, 15, and 20 $M_\odot$ 
from the Si core burning until the core collapse.
We showed for the first time detailed time variation of the neutrino emission relating the stellar 
evolution during the final stage.

The neutrino emission rate and the neutrino average energy increase during the Si core burning
and the collapsing stage.
However, they decrease during the O and Si shell burnings.
These nuclear burnings affect properties of the neutrino emission.
In these three models, larger stellar mass model indicates stronger neutrino emission and shorter period
from the Si burning to the core collapse.
Thus, the observations of preSN neutrinos could constrain the structure and burning processes 
in the central region of a preSN star.

Then, we estimated the neutrino events that will be observed by current and future neutrino detectors. 
After a massive star at $\sim 200$ pc forms an Fe core, 
the neutrino emission becomes so strong that a few to tens electron antineutrinos will be observed 
by KamLAND.
Future larger size scintillation detectors such as JUNO and RENO-50 could observe about 30 times as
large as neutrino events than KamLAND.
Hyper-Kamiokande could also observe hundreds neutrino events if
low observable threshold for preSN neutrinos is achieved.
We proposed for the first time a possibility of the observations of preSN neutrinos by Gd-loaded
Hyper-Kamiokande using delayed $\gamma$-ray signals from the neutron capture of Gd.
Although there are still many uncertainties in the threshold energy, detection efficiency, 
and the background events, Hyper-Kamiokande has a potential of observing thousands preSN
neutrino events owing to the large fiducial mass and the energy range of the delayed $\gamma$-ray signals.

If more than several neutrino events are observed in a unit time and the event number exceeds
the background event number, detailed variation of the neutrino events could be observed.
The decrease in the neutrino event rate will indicate the ignition of the O or Si shell burning.
These neutrino observations will be the first direct observation of the central region of
a fully evolved star.

SN alarm using preSN neutrino events is possible by KamLAND, SNO+, JUNO, and 
RENO-50 in one day before the explosion.
The alarm time will strongly depend on background events, especially on reactor neutrino experiments.
Monitoring transient neutrino events by multiple neutrino observatories raises the reliability of the SN
alarm.

Betelgeuse is a red supergiant with the initial mass of 13--20 $M_\odot$.
The distance is evaluated as $197 \pm 45$ pc.
Although the uncertainty in the distance is large, neutrino events could be observed by KamLAND, 
SNO+, JUNO, and RENO-50 for at most tens hours before the SN explosion.
The time variation of the neutrino events per hour might be observed by JUNO and RENO-50 
if the neutrino mass hierarchy is normal and the background by reactors is low.
Hyper-Kamiokande with Gd could observe thousands neutrino events using the detection of
delayed signals.
This observation could reveal burning processes in the central region of Betelgeuse.

\begin{acknowledgments}
We thank Chinami Kato and Shoichi Yamada for valuable discussions.
We are grateful to Masayuki Nakahata, Yusuke Koshio, and Hajime Yano 
for valuable comments about neutrino experiments.
This work has been partly supported by Grant-in-Aid for Scientific Research on Innovative Areas 
(26104007) from the Ministry of Education, Culture, Sports, Science and Technology (MEXT) and 
Grants-in-Aid for Scientific Research (24244028, 26400271) from Japan Society for the Promotion 
of Science.
\end{acknowledgments}

\nocite{*}

%\bibliography{ms.bib}% Produces the bibliography via BibTeX.


\end{document}